\documentclass[preprint,authoryear,11pt,letterpaper]{elsarticle}
\usepackage[top=0.75in, bottom=0.75in, left=0.75in, right=0.75in]{geometry}

\usepackage{amsmath,amsthm}
\usepackage{amssymb}
\usepackage{amsfonts}

\usepackage{graphicx}
\graphicspath{{graphics/}}

\usepackage[colorlinks]{hyperref}
\hypersetup{colorlinks,breaklinks,linkcolor=darkblue,urlcolor=darkblue,anchorcolor=darkblue,citecolor=darkblue}
\usepackage{booktabs}
\usepackage{multirow}
\usepackage{lscape}
\usepackage{subcaption}
\usepackage{epstopdf}
\usepackage{lineno}
\usepackage{caption}
\usepackage{enumitem}

\newcommand{\tensor}[1]{\boldsymbol{\mathcal{#1}}}
\newcommand{\mat}[1]{\boldsymbol{#1}}
\newcommand{\circshift}{\operatorname{circshift}}
\newcommand{\diag}{\operatorname{diag}}
\newcommand{\imag}{\operatorname{imag}}

\usepackage{booktabs}
\usepackage{multirow}

\usepackage{float}
\usepackage{microtype}
\usepackage{placeins}
\usepackage{xcolor}
\definecolor{deepblue}{rgb}{0,0,255}

\usepackage{algorithm2e}

\journal{Arxiv}

\begin{document}
	
	\begin{frontmatter}

		\title{Anti-circulant dynamic mode decomposition with sparsity-promoting for highway traffic dynamics analysis}

		\author[]{Xudong Wang}
		\ead{xudong.wang2@mail.mcgill.ca}

		\author[]{Lijun Sun\corref{cor1}}
		\ead{lijun.sun@mcgill.ca}

		\address{Department of Civil Engineering, McGill University, Montreal, QC H3A 0C3, Canada}

		\cortext[cor1]{Corresponding author. Address: 492-817 Sherbrooke Street West, Macdonald Engineering Building, Montreal, Quebec H3A 0C3, Canada}

\begin{abstract}
Highway traffic states data collected from a network of sensors can be considered a high-dimensional nonlinear dynamical system. In this paper, we develop a novel data-driven method---anti-circulant dynamic mode decomposition with sparsity-promoting (circDMDsp)---to study the dynamics of highway traffic speed data. Particularly, circDMDsp addresses several issues that hinder the application of existing DMD models: limited spatial dimension,  presence of both recurrent and non-recurrent patterns, high level of noise, and known mode stability. The proposed circDMDsp framework allows us to numerically extract spatial-temporal coherent structures with physical meanings/interpretations: the dynamic modes reflect coherent spatial basis, and the corresponding temporal patterns capture the temporal oscillation/evolution of these dynamic modes. Our result based on Seattle highway loop detector data showcases that traffic speed data is governed by a set of periodic components, e.g., mean pattern, daily pattern, and weekly pattern, and each of them has a unique spatial structure. The spatiotemporal patterns can also be used to recover/denoise observed data and predict future values at any timestamp by extrapolating the temporal Vandermonde matrix. Our experiments also demonstrate that the proposed circDMDsp framework is more accurate and robust in data reconstruction and prediction than other DMD-based models.

    \end{abstract}

		
\end{frontmatter}
	

\section{Introduction}

Spatiotemporal traffic state data collected from sensors on a highway network can be modeled as a high-dimensional nonlinear dynamical system, which mathematically characterizes how traffic variables (e.g., speed, volume, and density) evolve over time.
{A deep understanding of traffic dynamics behind the traffic sensing data provides valuable insights into the underlying complex traffic phenomena and, more importantly, enables both short-term and long-term predictions of future traffic states, which play a crucial role in supporting many applications in intelligent transportation systems (ITS). }

The abundance of large-scale traffic data has motivated researchers to develop data-driven models to reveal traffic dynamics and make predictions. One promising approach is dynamic mode decomposition (DMD), which was initially proposed in the field of fluid dynamics  \citep{schmid2010dynamic}. DMD depends on the fact that the Koopman operator \citep{koopman1932dynamical}---an infinite-dimension but linear operator that accurately describes a nonlinear dynamical system---can be well approximated by a large number of finite observables. It is an equation-free method that can extract coherent spatiotemporal structures from dynamical systems, {by decomposing the high-dimensional spatiotemporal data into a triplet of dynamic modes (spatial patterns), scalar amplitudes, and time dynamics (temporal patterns) \citep{schmid2022dynamic}. Unlike other matrix decomposition methods such as SVD, DMD provides a data-driven spectral decomposition, and the resulting spatiotemporal patterns identified by DMD have clear physical meanings---the dynamic modes reveal the structure of spatial basis, and the time dynamics specify the oscillation patterns of the corresponding dynamic modes \citep{brunton2016koopman}. The coherent spatiotemporal structure provides an analytical solution that allows us to perform data reconstruction, system diagnostics, and long-term extrapolation/prediction.}

DMD and its variants have achieved great success in analyzing spatiotemporal data from diverse domains beyond fluid dynamics, such as neural recordings, high-resolution video, and environmental data, to name but a few \citep{brunton2016extracting, kutz2016multiresolution, erichson2019compressed}. However, there are several critical challenges that hinder the direct application of DMD to analyze spatiotemporal traffic data. Firstly, a common prerequisite for DMD is that the spatial dimension should be much larger than the temporal dimension in order to find the best linear dynamical system approximation. However, given the fact that traffic data is constantly collected from a limited number of sensors  (e.g., at a scale of 100 to 1000) over a long time period, the true dynamics is often highly nonlinear and too complex to be approximated by a linear dynamical system. { Secondly, real-world traffic state data contains both recurrent patterns (e.g., morning rush-hour congestion) and nonrecurrent patterns (e.g., congestion caused by an unexpected incident/event) \citep{li2015trend}. In general, the recurrent/periodic data can be captured by low-frequency patterns with long-range periodic correlations, while those
anomalous and nonrecurrent data can be considered high-frequency variations (i.e., noise and pulse signals) with only short-range correlations. As DMD is mainly developed to extract coherent spatiotemporal patterns, the presence of nonrecurrent data could severely contaminate the estimated patterns in DMD. Thirdly, traffic systems are governed by daily and weekly periodicity due to the inherent rhythms in human mobility. The universal seasonality/periodicity in traffic data gives a strong prior and should serve as a constraint---we do expect the detect modes to be exactly stable (i.e., neither growing nor decaying). However, existing DMD models cannot ensure mode stability.}

In the literature, only a few studies have analyzed traffic dynamics using DMD-based models. \cite{liu2016data} was the first to use DMD to model highway traffic probe data, and the results reveal the heterogeneity of traffic dynamics in different time windows (e.g., morning peak hours/evening peak hours). \cite{avila2020data} utilized Hankel-DMD \citep{brunton2016extracting,arbabi2017ergodic} to analyze multi-lane highway traffic dynamics.  \citet{lehmberg2021modeling} studied crowd dynamics using an extended DMD \citep{williams2015data} with time delay embedding and the diffusion map algorithm. \cite{wang2022extracting} used Hankel-DMD and hierarchical clustering to study dynamic mobility patterns in the Guangzhou metro system.  All the above studies address the first challenge of insufficient spatial dimension through ``Hankelization'', which enlarges the spatial dimension of data by delay embedding.  These studies have shown that Hankel-DMD can indeed offer better interpretability and provide more accurate reconstruction and prediction of the data. However, Hankel-DMD still cannot encode the strong periodicity as a prior, and it remains susceptible to non-recurrent observations and outliers in the analyzed data, as is the case with traffic state data collected from field sensors.

{In this study, we propose a novel anti-Circulant Dynamic Mode Decomposition with Sparsity-promoting (circDMDsp) framework to encourage ``robustness'' and ``mode stability'' for spatiotemporal datasets that are predominantly recurrent but also contain considerable noise, outliers, and nonrecurrent observations. The circDMDsp framework incorporates two key features/designs. Firstly, we proposed to use an anti-circulant matrix instead of a Hankel matrix in expanding the spatial dimension. This augmentation is closely connected to the discrete Fourier transform (DFT)---the DFT matrix is a specific Vandermonde matrix where eigenvalues are roots of unity, so the circulant augmentation gives an effective solution to encourage mode stability. Secondly, we introduce a sparsity-promoting strategy to select the dominant patterns from the large set of candidate patterns; in addition, optimal hard threshold \citep{gavish2014optimal} is introduced to determine the optimal matrix rank, which allows the proposed model to be easily generalized to different traffic networks without tuning many parameters. The proposed circDMDsp framework is specifically designed for systems with strong seasonality, making it a suitable fit for analyzing highway traffic state data. The objectives of this research are three-fold: (1) discovering dynamical traffic patterns (Section 4.3); (2) reconstructing historical traffic speed and predicting future values (Section 4.2) based on the revealed spatiotemporal patterns; and (3) using DMD to quantify the short-term predictability of traffic state data at both sensor level and network level. }

The remainder of this paper is organized as follows. Section~\ref{sec:background} introduces notations and  the principles of Koopman spectral analysis andD, which are the fundamentals of the proposed framework. Some related DMD variants are also introduced in this section. In Section~\ref{sec:model}, we introduce the proposed anti-circulant dynamic mode with sparsity-promoting and its implementation/applications. Section~\ref{sec:result} presents a real-world case study based on Seattle highway traffic data. We compare the reconstruction and prediction performance with baseline models and analyze the discovered spatiotemporal patterns. Besides, we discuss short-term predictability at both the sensor level and the network level. In Section~\ref{sec:conclusion}, we summarize the study and discuss future research directions.

\section{Preliminary}
\label{sec:background}

\subsection{Notations}

Throughout the paper, we use letters to denote scalars, e.g., $x \in \mathbb{R}$ or  $X \in \mathbb{R}$, boldface lowercase letters to denote vectors, e.g., $\boldsymbol{x}\in \mathbb{R}^{n}$, boldface capital letters to denote matrices, e.g., $\boldsymbol{X}\in \mathbb{R}^{n_1 \times n_2}$. We denote the $(i,j)$th entry of a matrix by $\mat{X}_{i,j}$. The Frobenius norm of $\mat{X}$ is defined as $\|\mat{X}\|_F = \sqrt{\sum_{i=1}^{n_1}\sum_{j=1}^{n_2} \mat{X}_{i,j}^2}$ and the $l_1$-norm of $\|\mat{x}\|_1 =\sum_{i=1}^n |x_i|$.

\subsection{Koopman analysis}

This section provides a brief introduction to the fundamental theory of Koopman spectral analysis, which forms the basis of DMD. We refer readers to \cite{kutz2016dynamic} and  \cite{schmid2022dynamic} for more details about the theory of the Koopman analysis and its connection with DMD. In a \textit{nonlinear} discrete-time dynamical system, a state $\mat{x}_t \in \mathbb{R}^N$ in state space $\mathcal{M}$ at time $t$ can be formulated as:
\begin{equation}
\label{eq:nonlinear}
    \mat{x}_{t+1} = \mat{f}(\mat{x}_t),
\end{equation}
where $\mat{f}(\cdot)$ describes the nonlinear dynamics of the system, e.g., ordinary differential equations of states, mapping the state $\mat{x}_t$ over a given time interval $\Delta t$ onto $\mat{x}_{t+1}$.

Traditional ways to analyze \eqref{eq:nonlinear} are through investigating equilibrium points and the linear dynamics of small perturbations, if $\mat{f}$ can be formulated via conservation laws and physical models \citep{schmid2022dynamic}. However, in real-world systems, the dynamics $\mat{f}$ are often too difficult or impractical to formulate. Koopman analysis is designed to extract pivotal coherent structures from data without knowing $\mat{f}$. Rather than based on state variable $\mat{x}_t$, Koopman analysis rests on the observables of state $\mat{g}(\mat{x}_t): \mathcal{M} \rightarrow \mathbb{C}$ in an observable space $\mathcal{G}$. An \textit{infinite-dimensional linear operator} called Koopman operator $\mathcal{K}$ acts on the observable function $\mat{g}$ such that $\mathcal{K}\mat{g} = \mat{g} \circ \mat{f}$, where $\circ$ denotes the composition of $\mat{g}$ with $\mat{f}$ \citep{brunton2016koopman}. With the Koopman operator $\mathcal{K}$, the analysis of nonlinear dynamics on $\mat{x}_t$ in \eqref{eq:nonlinear} equals to an infinite-dimensional but linear problem on $\mat{g}(\mat{x}_t)$:
\begin{equation}
\label{eq:koopman}
    \mathcal{K}\mat{g}(\mat{x}_t) = \mat{g}(\mat{f}(\mat{x}_t))=\mat{g}(\mat{x}_{t+1}).
\end{equation}

\begin{figure}[!ht]
    \centering
    \includegraphics{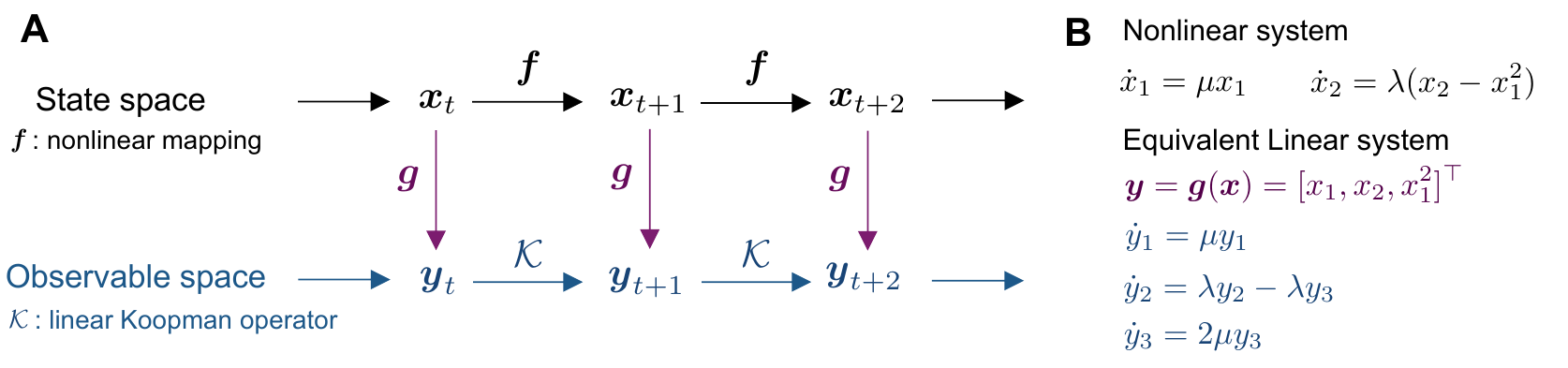}
    \caption{A: Illustration of the Koopman operator for nonlinear dynamical system; B: an example of applying Koopman operator on a nonlinear dynamical system.}
    \label{fig:Koopman}
\end{figure}

Figure \ref{fig:Koopman} illustrates the procedure of transforming a nonlinear system in the state variables into a linear system in the observables via observable functions $\mat{g}(\mat{x}_t)$ and explains the key concepts of Koopman analysis by a simple nonlinear dynamical system. Given suitable observables (e.g., in Figure \ref{fig:Koopman}B), the nonlinear dynamical system can be equivalently described as a linear system.

To determine such observables, the invariant subspace of $\mathcal{K}$ is considered by eigendecomposition of $\mathcal{K}$:
\begin{equation}
\label{eq:eigen}
    \mathcal{K} \varphi_k(\mat{x}) = \lambda_k \varphi_k(\mat{x}), \quad k \in \mathbb{N},
\end{equation}
where $\lambda_k \in \mathbb{C}$ is the $k$-th eigenvalue (\textit{Koopman eigenvalue}) and $\varphi_k:\mathcal{M}\rightarrow \mathbb{C}$ is the corresponding eigenfunction (\textit{Koopman eigenfunction}). The Koopman eigenfunctions can be regarded as a set of basis functions (intrinsic measurement coordinates); if $\mat{g}$ lies within the span of the eigenfunctions, the observables can be expressed as a linear combination of those eigenfunctions, e.g., $\mat{g}(\mat{x}_t) = \sum_{k=1}^{\infty} {c}_k \varphi_k(\mat{x}_t)$. In practice, it is challenging to obtain the eigenfunctions and eigenvalues from the infinite-dimensional Koopman operator $\mathcal{K}$.

\subsection{Dynamic mode decomposition and its variants}

DMD is a data-driven method to approximate the eigenvalues and eigenvectors of the infinite Koopman operator when the number of observables (spatial dimension of data) tends to infinity \citep{korda2018convergence}. DMD aims to find these leading eigenvectors $\mat{\Phi}$ and eigenvalues $\mat{\lambda}$ of the best-fit linear operator $\mat{A}$ for all the data $\mat{X} \in \mathbb{R}^{N \times T}$ by solving a least square problem:
\begin{equation}
\label{eq:xax}
    \underset{\mat{A}}{\min} \|\mat{X}_2 -\mat{AX}_1\|^2_F.
\end{equation}
where $\mat{X}_1=[\mat{x}_1, \cdots, \mat{x}_{T-1}]\in \mathbb{R}^{N \times (T-1)}$, $\mat{X}_2=[\mat{x}_2, \cdots, \mat{x}_{T}]\in \mathbb{R}^{N \times (T-1)}$. In dynamic systems, the eigenvectors (dynamic modes) of dynamic matrix reflect the spatial correlations, and the corresponding eigenvalues determine the time dynamics---including oscillation frequency and decay/growth rate---of these modes. {In practice, the state dimension $N$ is large, it is intractable to directly solve the optimal problem \eqref{eq:xax}. To overcome this challenge, DMD analyzes a low-rank representation of $\mat{A}$, obtained through a proper orthogonal decomposition (POD) projection, represented as $\mat{\Tilde{A}}$, instead of performing a direct eigendecomposition of $\mat{A}$. Algorithm \ref{alg:proposed} summarizes the procedure of DMD and future details can be found in Section \ref{sec:circdmd}. }

\begin{algorithm}[htbp]
\caption{Dynamic Mode Decomposition (DMD)}\label{alg:proposed}
\KwData{$\mat{X} \in \mathbb{R}^{N \times T}$}
\KwResult{$\mat{\Phi}\in\mathbb{C}^{N\times r}$, $\mat{b}\in\mathbb{C}^{r\times r}$, $\mat{\Psi}\in\mathbb{C}^{r\times T}$}
\text{Step 1: Define matrices $\mat{X}_1$ and $\mat{X}_2$ from the data $\mat{X}$};

\text{Step 2: Take the reduced SVD of matrix $\mat{X}_1$} to compute $\mat{U}_r, \mat{\Sigma}_r$ and $\mat{V}_r$ with rank $r$, i.e., $\mat{X}_1\approx \mat{U}_r \mat{\Sigma}_r \mat{V}_r^\star$;

\text{Step 3: Project $\mat{A}$} to POD subspace spanned by $\mat{U}_r$ to obtain the low-rank dynamical matrix $\mat{\Tilde{A}}$: $\mat{\Tilde{A}} = \mat{U}_r^\top\mat{A}\mat{U}_r = \mat{U}_r^\top\mat{XV}_r\mat{\Sigma}_r^{-1}$;

\text{Step 4: Compute the eigendecomposition of $\mat{\Tilde{A}}$} to get eigenvectors (dynamic modes) $\mat{\Phi}$ and eigenvalues $\mat{\lambda}$: $  \mat{\Tilde{A}}\mat{W} = \mat{W} {\diag}(\mat{\lambda})$ and $\mat{\Phi} = \mat{X}_2\mat{V}_r\mat{\Sigma}_r^{-1}\mat{W}$;

\text{Step 5: Calculate the amplitude $\mat{b}$ in terms of initial condition $\mat{x}_1$ and dynamic modes $\mat{\Phi}$: $\mat{b} = \mat{\Phi}^\dagger \mat{x}_1$. }

\text{Step 6: Construct a Vandermonde matrix $\mat{\Psi}$ with eigenvalues of a geometric progression in each row.}
\end{algorithm}

{Similarly to singular value decomposition (SVD), where $\mat{X} = \mat{USV}^\top$, the data $\mat{X}$ is also decomposed into three low-rank matrices such that $\mat{X} = \mat{\Phi}\diag(\mat{b})\mat{\Psi}$ by DMD, where $\mat{\Psi}$ is a Vandermonde matrix with eigenvalues.  The main difference between DMD and SVD is that in SVD, the low-rank matrices $\mat{U}$ and $\mat{V}$ are orthogonal, resulting in each individual pattern being a mixture to represent the data. In contrast, DMD incorporates a dynamic constraint to promote the separation of the data into sets of oscillation frequencies and modes. The modes evolve in accordance with their respective oscillation frequencies. However, the elements $\mat{\lambda}$  of Vandermonde matrix $\mat{\Psi}$ are not necessarily roots of unity in DMD, which can result in unstable evolution. In particular, a mode with growing evolution will diverge to infinity as the window size increases, whereas a dynamic mode with decaying evolution will ultimately converge to zero.}

{DMD was initially introduced in the field of fluid flows, where the spatial dimension of fluid flows is typically much larger than the temporal dimension, i.e., $N \gg T$. However, in the context of transportation networks, it is common to have hundreds of sensors or stations collecting thousands of traffic variables per week. As a result, the resulting data matrix $\mat{X}$ over a study time window often has $N < T$. In this case, the true dynamics is highly nonlinear, and the low-rank approximation of the time evolution---represented by $\mat{\Psi}$ with only $r$ parameters---has limited capacity to model the complex data. Consequently, it becomes challenging to accurately approximate the dynamics in $\mat{X}$ using a simple linear dynamical system.} The most common way to enable linear approximation is through data augmentation. Hankel dynamic mode decomposition (HankelDMD) employs DMD on a Hankel transformation of data \citep{brunton2016extracting,avila2020data, lehmberg2021modeling,wang2021hankel}. {Given an $N \times T$ matrix $\mat{X}$, the Hankel matrix $\mat{H}\in\mathbb{R}^{N\tau \times (T-\tau+1)}$ with the delay embedding length $\tau$ is defined as
\begin{equation}
\label{eq:hankel}
\mat{H} =
  \begin{bmatrix}
        \mat{x}_1 & \mat{x}_2 & \dots & \mat{x}_{T-\tau+1}\\
        \mat{x}_2 & \mat{x}_3 & \dots & \mat{x}_{T-\tau+2}\\
        \vdots & \vdots & \ddots & \vdots \\
        \mat{x}_\tau & \mat{x}_{\tau+1} & \cdots & \mat{x}_{T}\\
    \end{bmatrix} = \begin{bmatrix}
        \mat{h}_1 & \dots & \mat{h}_{T-\tau+1}
    \end{bmatrix}.
\end{equation}
where the spatial dimension is increased from $N$ to $N\tau$. Then, one can perform the default DMD analysis on $\mat{H}$ instead of the original data $\mat{X}$.}

Another key issue in applying DMD is that real-world data generated from field sensor networks often contain considerable noises, outliers, and also nonrecurrent patterns (e.g., short-term traffic congestion caused by a minor event/incident), that can hardly be explained by the principle dynamics. The default DMD algorithm is very sensitive to noises and outliers; more important, the existence of nonrecurrent temporal data will lead to biased estimations of the temporal evolution in terms of amplitude and periodicity. To address this issue, debiased algorithms are proposed to correct the bias by defining $\mat{\Tilde{A}}$ as $\mat{A}$ in POD subspace. For example, forward-backward DMD (fbDMD) \citep{dawson2016characterizing} separately models forward dynamics $\mat{X}_2\approx \mat{A}_f\mat{X}_1$ and back dynamics $\mat{X}_1\approx \mat{A}_b\mat{X}_2$, with two different propagators:
\begin{equation}
    \mat{\Tilde{A}}_f = \mat{U}_{1}^\top\mat{X}_2\mat{V}_{1}\mat{\Sigma}_{1}^{-1}, \quad \mat{\Tilde{A}}_b = \mat{U}_{2}^\top\mat{X}_1\mat{V}_{2}\mat{\Sigma}_{2}^{-1},
\end{equation}
where $\mat{X}_1 \approx \mat{U}_1\mat{\Sigma}_1\mat{V}_1$ and  $\mat{X}_2 \approx \mat{U}_2\mat{\Sigma}_2\mat{V}_2$ are reduced singular value decompositions (SVD)s of $\mat{X}_1$ and $\mat{X}_2$, respectively. Then a debiased estimation of the forward propagator is given by
\begin{equation}
    \mat{\Tilde{A}} = \left(\mat{\Tilde{A}}_f \mat{\Tilde{A}}_b^{-1}\right)^{1/2}.
\end{equation}
Due to the nonuniqueness of the square root, the eigenvalues of $\mat{\Tilde{A}}$ are determined up to a factor of $\pm 1$ by the square root \citep{askham2018variable}.
{Another well-known debiased DMD is \textit{total-least-square} DMD (tlsDMD) \citep{hemati2017biasing}, which aims to seek an exact linear relationship between noise-free data in both $\mat{X}_1$ and $\mat{X}_2$:
\begin{equation}
\label{eq:tdmd}
\begin{aligned}
\underset{\mat{A},\Delta\mat{X}_1,\Delta\mat{X}_2}{\min } \ & \bigg{\Vert}\begin{bmatrix}
    \Delta \mat{X}_1 \\ \Delta \mat{X}_2
    \end{bmatrix}\bigg{\Vert}_F \\
    \text{s.t. } \ &  \mat{X}_2 - \Delta \mat{X}_2 = \mat{A}(\mat{X}_1 - \Delta \mat{X}_1).
\end{aligned}
\end{equation}
The optimization problem \eqref{eq:tdmd} can be solved by projecting $\mat{X}_1$ and $\mat{X}_2$ onto POD modes spanned by an augmented matrix
\begin{equation}
    \mat{Z} =
  \begin{pmatrix}
        \mat{X}_1 \\
        \mat{X}_2
    \end{pmatrix}.
\end{equation}
The reduced SVD is calculated of $\mat{Z}$, $\mat{Z} \approx \mat{U}_z \mat{\Sigma}_z \mat{V}_z^\ast$ with the first $z$ eigenvalues. The projected matrices
\begin{equation}
\mat{\Bar{X}}_1=\mat{X}_1\mathbb{P}_{\mat{Z}^\ast},\quad \mat{\Bar{X}}_2=\mat{X}_2\mathbb{P}_{\mat{Z}^\ast},
\end{equation}
where $\mathbb{P}_{\mat{Z}^\ast} = \mat{V}_z\mat{V}_z^\ast$, are used to replace the original data in DMD procedure.}

\section{Anti-circulant dynamic mode decomposition with sparsity-promoting (circDMDsp)}
\label{sec:model}

\begin{figure}[!ht]
    \centering
\includegraphics{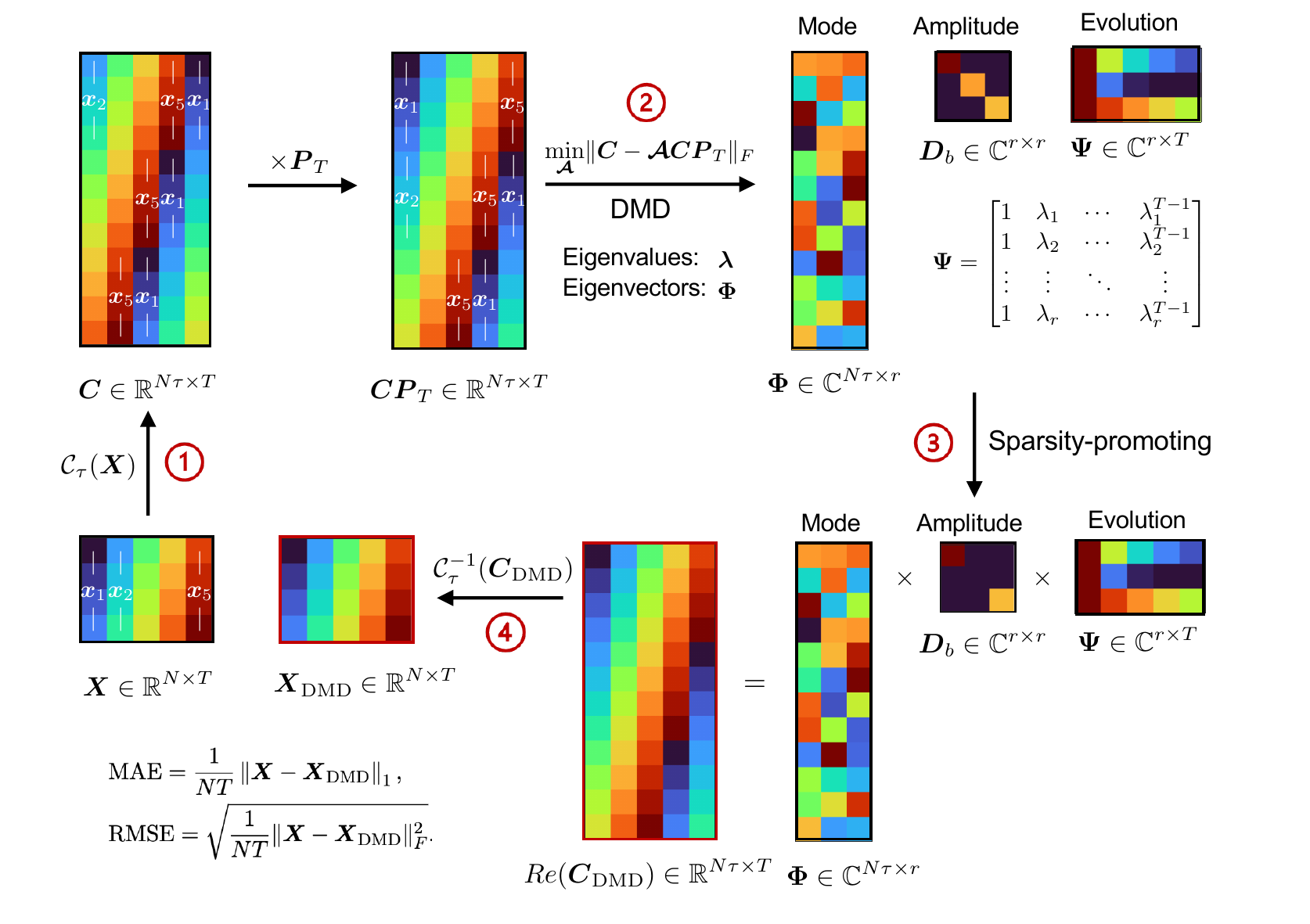}
    \caption{{An illustration of the proposed circDMDsp when $\tau = 3$ for a $4\times 5$ matrix $\mat{X}$. The model includes four steps: (1) applying anti-circulant operator to enlarge the number of observables; (2) using DMD to decompose the anti-circulant matrix into dynamic modes, amplitude and temporal evolution; (3) employing a sparsity-promoting method (DMDSP) to obtain a concise model; (4) utilizing the anti-circulant structure to conduct historical data reconstruction and future values prediction. }}
    \label{fig:dmdflow}
\end{figure}

{As previously mentioned, the Vandermonde matrix in DMD may result in modes with divergent (growing) or convergent (decaying) evolution, negatively impacting the accuracy of long-term traffic prediction. To mitigate this issue, we propose an anti-circulant matrix that is designed to have a desirable connection with the Discrete Fourier Transform (DFT), in which the DFT matrix is a specific type of Vandermonde matrix where the elements are chosen to be roots of unity. The proposed anti-circulant dynamic mode decomposition (circDMD) leverages the fact that the anti-circulant matrix can be diagonalized by DFT.} The framework consists of four key steps as outlined in Figure \ref{fig:dmdflow}. Firstly, we organize the raw traffic data as a matrix $\mat{X}$ (sensor $\times$ time point) and then transform it into an anti-circulant matrix $\mat{C}$ using a circulant data augmentation operator $\mathcal{C}_\tau$.  Secondly, we assume the augmented data follows a locally linear system at any timestamp $t$ such that $\mat{c}_{t+1} \approx \tensor{A}\mat{c}_t$, where $\tensor{A}$ is an $N\tau \times N\tau$ dynamic matrix. We estimate the optimal dynamic matrix $\tensor{A}$ using DMD, which provides the dynamic modes $\mat{\Phi}$, temporal evolution $\mat{\Psi}$ and corresponding amplitude $\mat{D}_b$.
Thirdly, we apply a sparsity-promoting DMD method (DMDSP) \citep{jovanovic2014sparsity} to find a concise representation by setting the amplitudes of insignificant modes to zero. Based on the refined amplitudes, we then select the dominant modes and use these modes to analyze the dynamics of the traffic network, reconstruct historical data, and predict future values. Finally, we use an inverse anti-circulant operator $\mathcal{C}_\tau^{-1}$ to restore the estimation (reconstruction and prediction) in the anti-circulant space to the original space. As our model ensures unitary roots, it provides a natural solution for periodic data and is more robust to outliers data.

\subsection{Anti-circulant dynamic mode decomposition (circDMD)}
\label{sec:circdmd}

{First of all, let define an operator $\circshift(\mat{X}, L)$ that circularly shift each column of $\mat{X}$ by $L$ position to the right ($L$ is negative) or to the left ($L$ is positive). Then, an anti-circulant matrix $\mat{C}$ constructed by an anti-circulant operator $\mathcal{C}_\tau(\mat{X}):\mathbb{R}^{N \times T} \Rightarrow \mathbb{R}^{N\tau \times T}$ on $\mat{X}$ for $ i=1,\dots, \tau$ is:
\begin{equation}
\label{eq:circ}
    \mathcal{C}_\tau(\mat{X}) := \mat{C}_{(i-1)N+1:iN,~ :} = {\circshift}(\mat{X},-i) =   \begin{bmatrix}
\mat{x}_2& \mat{x}_3 & \cdots & \mat{x}_T  & \mat{x}_1\\
\mat{x}_3& \mat{x}_4 & \cdots & \mat{x}_1 & \mat{x}_2\\
\vdots& \vdots&\ddots& \vdots & \vdots \\
\mat{x}_{\tau+1}& \mat{x}_{\tau+2} & \cdots & \mat{x}_{\tau-1} & \mat{x}_\tau\end{bmatrix}=\begin{bmatrix}
    \mat{c}_2, \mat{c}_3, \ldots, \mat{c}_T,\mat{c}_1
\end{bmatrix}.
\end{equation}
Let a $T$-by $T$ permutation matrix be $\mat{P}_T = \mat{I}_T(:,p), ~p=\left[T,~ 1:T-1\right]$, we have
 \begin{equation}
     \mat{CP}_T = \begin{bmatrix}
\mat{x}_1& \mat{x}_2 & \cdots & \mat{x}_{T-1} & \mat{x}_T  \\
\mat{x}_2& \mat{x}_3 & \cdots & \mat{x}_T &\mat{x}_1\\
\vdots& \vdots&\ddots& \vdots & \vdots \\
\mat{x}_\tau& \mat{x}_{\tau+1} & \cdots & \mat{x}_{\tau-2} & \mat{x}_{\tau-1}\end{bmatrix}=\begin{bmatrix}
    \mat{c}_1 & \dots &\mat{c}_T
\end{bmatrix}.
\label{eq:anti-circ}
 \end{equation}
As we can see, the construction of the anti-circulant matrix is very similar to ``Hankelization'' in Eq.~\eqref{eq:hankel}. However, it is important to note that the anti-circulant matrix described in Equation \eqref{eq:anti-circ} is a circular data structure, which can be diagonalized using the Discrete Fourier Transform (DFT). When $\tau=T$, the result of $\mathcal{C}_T(\mat{X})\mat{P}_T$ becomes a full block anti-circulant matrix and we have \citep{yamamoto2022fast}
\begin{equation}
\label{eq:flf}
   \mathcal{C}_T(\mat{X})\mat{P}_T =
        \left(\mat{I}_N \otimes \mat{F}_T\right)\begin{bmatrix}
        \mat{\Theta}_1\\
        \mat{\Theta}_2\\
        \vdots\\
        \mat{\Theta}_N
    \end{bmatrix}\mat{F}_T,
\end{equation}
where $\mat{F}_T\in \mathbb{C}^{T \times T}$ is discrete Fourier matrix, $\mat{F}_T^*\in \mathbb{C}^{T \times T}$ is conjugate discrete Fourier matrix, $\mat{\Theta}_i = \diag(\frac{1}{T}\mat{F}_T^* \mat{X}_{i,:}) \in \mathbb{C}^{T\times T}$ for the $i$th row of $\mat{X}$. Let $\mat{Q}_\tau = [\mat{I}_\tau, ~ \mat{0}_{\tau, T-\tau} ]\in \mathbb{R}^{\tau\times T}$ be a truncation matrix, then for delay embedding length $\tau$, we have
\begin{equation}
\label{eq:flf2}
   \mathcal{C}_\tau(\mat{X})\mat{P}_T =
        \left(\mat{I}_N \otimes (\mat{Q}_\tau\mat{F}_T)\right)\begin{bmatrix}
        \mat{\Theta}_1\\
        \mat{\Theta}_2\\
        \vdots\\
        \mat{\Theta}_N
    \end{bmatrix}\mat{F}_T,
\end{equation}
Eq.~\eqref{eq:flf2} indicates that the right part of $\mat{CP}_T$ is a discrete Fourier matrix where eigenvalues are roots of unity, eliminating the varying modes. Therefore, the additional data makes the model more robust in characterizing key dynamic modes that produce stable oscillations that match the underlying daily/weekly periodicity of traffic state data.
 }

In anti-circulant dynamic mode decomposition (cricDMD), we assume the traffic system is a locally linear dynamical system so that $\mat{c}_{t+1} \approx \tensor{A}\mat{c}_t$, where $\mat{c}_t \in \mathbb{R}^{N\tau}$ is the $t$ column of $\mat{CP}_T$ and $\tensor{A}\in \mathbb{R}^{N\tau \times N\tau}$ is \textit{dynamic matrix}. Then, the objective of circDMD turns to find the leading eigendecomposition of $\tensor{A}$ relating to $\mat{CP}_T$ and $\mat{C}$ by solving
\begin{equation}
\label{eq:cac}
    \underset{\tensor{A}}{\min} \ \|\mat{C} -\tensor{A}\mat{CP}_T\|_F.
\end{equation}
and the solution is:
\begin{equation}
\label{eq:dmd}
    \tensor{A}=\mat{C}(\mat{CP}_T)^{\dagger},
\end{equation}
where ${\dagger}$ means Moore–Penrose inverse, which can be calculated by SVD, i.e., $(\mat{CP}_T)^\dagger \approx \mat{V}\mat{\Sigma}^{-1}\mat{U}^\top$. However, it is intractable to analyze $\tensor{A}$ directly from Eq. \eqref{eq:dmd} when the dimension $N\tau$ or/and $T$ is large. DMD was proposed to tackle the dimension problem by projecting $\tensor{A}$ to a lower-dimension space spanned by first $r$ left singular vectors $\mat{U}_r \in \mathbb{R}^{N\tau \times r}$ \citep{tu2013dynamic}, i.e., POD subspace:
\begin{equation}
\label{eq:atilde}
    {\tensor{\Tilde{A}}} = \mat{U}_r^\top\tensor{A}\mat{U}_r=\mat{U}_r^\top\mat{C}\mat{V}_r\mat{\Sigma}_r^{-1},
\end{equation}
where $\mat{\Sigma}_r \in \mathbb{R}^{r \times r}$ and $\mat{V}_r \in \mathbb{R}^{T \times r}$, and $r$ denotes the rank of $\tensor{\Tilde{A}}$.

{The matrices $\tensor{A}$ and $\tensor{\Tilde{A}}$ are related via similarity transform and they have the same nonzero leading eigenvalues}, so the eigenvalues $\mat{\lambda} \in \mathbb{C}^{r}$ of $\tensor{A}$ can be calculated from $\tensor{\Tilde{A}}$:
\begin{equation}
\label{eq:eigen}
    \tensor{\Tilde{A}}\mat{W} = \mat{W} {\diag}(\mat{\lambda}),
\end{equation}
where the columns of $\mat{W}\in \mathbb{C}^{r\times r}$ are the eigenvectors of $\tensor{\Tilde{A}}$, then the eigenvectors $\mat{\Phi} \in \mathbb{C}^{N\tau \times r}$ of $\tensor{A}$ is described as \citep{kutz2016dynamic}:
\begin{equation}
\label{eq:phi}
    \mat{\Phi} := \mat{C}\mat{V}_r\mat{\Sigma}_r^{-1}\mat{W} \quad (\text{exact dynamic modes}) \quad \text{or} \quad \mat{\Phi}:=\mat{U}_r\mat{W} \quad (\text{projected dynamic modes}),
\end{equation}
where these two definition will tend to converge if $\mat{C}$ and $\mat{CP}_T$ have the same column spaces \citep{kutz2016dynamic}.

The appeal of DMD lies in its ability to interpret the physical phenomena underlying the data through the eigenvalues and eigenvectors of the dynamic matrix. The dynamic modes (i.e., eigenvectors) $\mat{\Phi}$ depict the spatially coherent  structures in the data, and the corresponding eigenvalues define growth/decay rates and oscillation frequencies for each mode \citep{tu2013dynamic}. For each eigenpair $(\phi_i,~ \lambda_i)$, the magnitude of the eigenvalue $|\lambda_i|$ determines whether the corresponding dynamic mode $\phi_i$ is steady (when $|\lambda_i|=1$) or varying (decaying if $|\lambda_i|<1$, and growing if $|\lambda_i|>1$), and the imaginary part of $\lambda_i$ determines the oscillation period of $\phi_i$. As a result, DMD incorporates spatial dimensionality-reduction techniques with Fourier transforms in the time domain \citep{kutz2016dynamic}.

With the detected eigenvalues $\lambda_i$ and the eigenvectors $\phi_i$ of $\tensor{A}$, the state variable $\mat{c}_t$ at any timestamp $t$ can be approximated as
\begin{equation}
\label{eq.ct}
    \mat{c}_t \approx \sum_{i=1}^r b_i \mat{\phi}_i \lambda_i^{t-1}
\end{equation}
{where $\mat{b}$ is the mode amplitudes computed as $\mat{b} = \mat{\Phi}^{\dagger}\mat{c}_1$} with  the initial condition $\mat{c}_1=\mat{C}(:,T)$. It is also possible to extrapolate values at any given time $t$ using the continues-time formula
\begin{equation}
\label{eq:extrapo}
    \mat{c}(t) \approx \sum_{i=1}^r b_i \mat{\phi}_i e^{\log(\lambda_i)(t-1)/\Delta t}.
\end{equation}

 We denote $\mat{C}_{\text{DMD}} \in \mathbb{C}^{N\tau \times T}$ as the recovered matrix by circDMD, and the expression \eqref{eq.ct} can be written in a matrix form
\begin{equation}
\label{eq:cdmd}
\mat{C}_\text{DMD} =
\begin{bmatrix}
    \mat{\phi}_1 & \mat{\phi}_2 & \dots & \mat{\phi}_{r}
\end{bmatrix}
\begin{bmatrix}
b_1 &  &  &  \\
& b_2 &  & \\
& & \ddots&  \\
& & & b_r\end{bmatrix}
\begin{bmatrix}
1& \lambda_1 & \cdots & \lambda_1^{T-1}  \\
1& \lambda_2 & \cdots & \lambda_2^{T-1}\\
\vdots& \vdots&\ddots& \vdots  \\
1& \lambda_r & \cdots & \lambda_r^{T-1}\end{bmatrix}=
\mat{\Phi}\mat{D}_b\mat{\Psi},
\end{equation}
where $\mat{D}_b$ is a diagonal matrix of amplitudes $\mat{b}$, and $\mat{\Psi} \in \mathbb{C}^{{r} \times T}$ is a Vandermonde matrix of eigenvalues denoted time dynamics. We can achieve system prediction ($T_\xi > T$) by adjusting the time window length $T_\xi$ in the Vandermonde matrix $\mat{\Psi}$. In practice, we use the real part of $\mat{C}_\text{DMD}$ as the estimation. As the reconstruction presented in Eq.~\eqref{eq:cdmd} is  in the delay embedding space, we still need to transform it back to the original space. Instead of using the first $N$ elements of each mode \citep{brunton2016extracting,avila2020data}, we utilize all $N\tau$ modes to conduct the transformation. We define an inverse anti-circulant operator $\mathcal{C}_\tau^{-1}(\mat{C}): \mathbb{R}^{N\tau \times T} \Rightarrow \mathbb{R}^{N \times T}$, i.e., $\mathcal{C}_\tau^{-1}(\mathcal{C}_\tau(\mat{X}))=\mat{X}$:
\begin{equation}
\label{eq:inv_circ}
    \mathcal{C}_\tau^{-1}(\mat{C}) := \sum_{i=1}^\tau {\circshift}\left(\mat{C}_{(i-1)N+1:iN,~ :}, i-1\right)/\tau.
\end{equation}

\subsection{Refining DMD with sparsity-promoting}

In many cases, the rank $r$ would be large, resulting in a verbose model representation. We utilize a sparsity-promoting DMD (DMDSP) method \citep{jovanovic2014sparsity} to further obtain a concise expression based on the result from circDMD. Our goal is to select only the most important patterns, and this can be achieved by  the following optimization problem:
\begin{equation}
\label{eq:sp}
    \underset{\mat{b}}{\min} \ \|\mat{CP}_T - \mat{\Phi}\mat{D}_b\mat{\Psi}\|_F + \gamma\|\mat{b}\|_1,
\end{equation}
{where $\gamma$ is a parameter chosen to control the sparsity of the mode amplitudes $\mat{b}$---a larger value of $\gamma$ will give more zeros terms in $\mat{b}$. The DMDSP algorithm includes two steps:
\begin{enumerate}[label=(\arabic*)]\setlength\itemsep{0em}
    \item Determining the sparsity structure, which balances the approximation error and the amplitudes of modes by alternating direction method of multipliers (ADMM) framework (i.e., Problem \eqref{eq:sp})
    \item Determining the amplitudes of the selected modes under the sparsity structure by solving an equality-constrained quadratic programming problem.
\end{enumerate}
Note that the projected dynamic modes are used in DMDSP. The details of DMDSP can be found in \cite{jovanovic2014sparsity}.  }

\subsection{Implementation}

Three hyper-parameters in the proposed model need to be determined: delay embedding length $\tau$ in the anti-circulant matrix, rank $r$ of the reduced matrix in DMD, and  trade-off parameter $\gamma$ in DMDSP. {Recall that the delay embedding matrix $\mat{C}$ of data $\mat{X}$ is an $N\tau \times T$ matrix where the spatial dimension $N\tau$ should be much larger than the temporal dimension $T$ to meet the Koopman condition. We set the delay embedding length to be the period length of data, e.g., daily timestamps for traffic data, to capture the periodicity. Especially, the matrix $\mat{C}$ would be a fully block anti-circulant matrix when $\tau = T$. Larger $\tau$ can increase the performance of circDMDsp, but it decreases computational efficiency since it uses $\mat{CP}_T(\mat{CP}_T)^\top \in \mathbb{R}^{N\tau \times N\tau}$ to obtain left singular vectors $\mat{U}$ in SVD. To overcome the curse of dimensionality, we leverage the method of snapshots to compute a much smaller matrix $(\mat{CP}_T)^\top\mat{CP}_T \in \mathbb{R}^{T \times T}$ to obtain the right singular vectors $\mat{V}$ at first. The size of the shorter matrix is irrelevant to the delay embedding length $\tau$. The left singular vectors $\mat{U}$ in \eqref{eq:atilde} can be approximated by $\mat{CP}_T\mat{V}\mat{\Sigma}^{-1}$ after obtaining $\mat{V}$ and $\mat{\Sigma}$. For the matrix rank $r$ in DMD, we apply optimal hard threshold \citep{gavish2014optimal} $\delta$ to determine it:
\begin{equation}
\label{eq:threshold}
    \delta = (0.56\beta^3-0.95\beta^2+1.82\beta+1.43)\sigma_\text{med},
\end{equation}
where $\beta = T/N\tau$ and $\sigma_\text{med}$ is the median of singular values $\diag(\mat{\Sigma})$, and only the singular values larger than the threshold are kept. The sparsity parameter $\gamma$ controls the ultimate number of modes. In Section \ref{sec:reconstruction}, we set different values of $\gamma$ to examine the reconstruction results.}

To quantify the estimation performance, mean absolute error (MAE) and root mean square error (RMSE) are used for overall evaluation:
\begin{equation}
\begin{aligned}
\label{eq:rmse}
    &\text{MAE} = \frac{1}{NT_\xi}\left\|\mat{X}-{\mat{X}_\text{DMD}}\right\|_1,\\
    &\text{RMSE} = \sqrt{\frac{1}{NT_\xi}\|\mat{X}-{\mat{X}}_\text{DMD}\|_F^2}.
\end{aligned}
\end{equation}
For the predictability analysis in Section \ref{sec:predict}, we use mean absolute percentage error (MAPE) to evaluate the $n$th sensor:
\begin{equation}
\label{eq:mape}
        \text{MAPE}(n) = \frac{100\%}{T}\left\|\frac{\mat{X}(n,:)-{\mat{X}_\text{DMD}(n,:)}}{\mat{X}(n,:)}\right\|_1.
\end{equation}

\section{Case study}
\label{sec:result}

In this study, we employ the proposed circDMDsp framework to analyze real-world highway traffic speed data. The objective of this study is threefold: (1) to investigate the spatial and temporal characteristics of highway traffic state, (2) to reconstruct the historical traffic speed, and (3) to make predictions of future values based on the identified dynamic traffic patterns.

\subsection{Highway traffic speed dataset}
\label{sec:data}

We use traffic speed data collected from loop detectors on the highway networks of Seattle (Washington, USA) \citep{cui2019traffic} to evaluate the proposed framework. The sensors are installed on four highways: I-5, I-90, I-405, and SR-520. The data set includes 157 sensors on the southbound (SB) and 165 sensors on the northbound (NB), with a temporal resolution of 5 minutes (288 time points per day). For the purpose of evaluation, we employ the first two weeks of data for training and utilize the data from the subsequent one week to assess the prediction performance.

\begin{figure}[!h]
    \centering
    \includegraphics{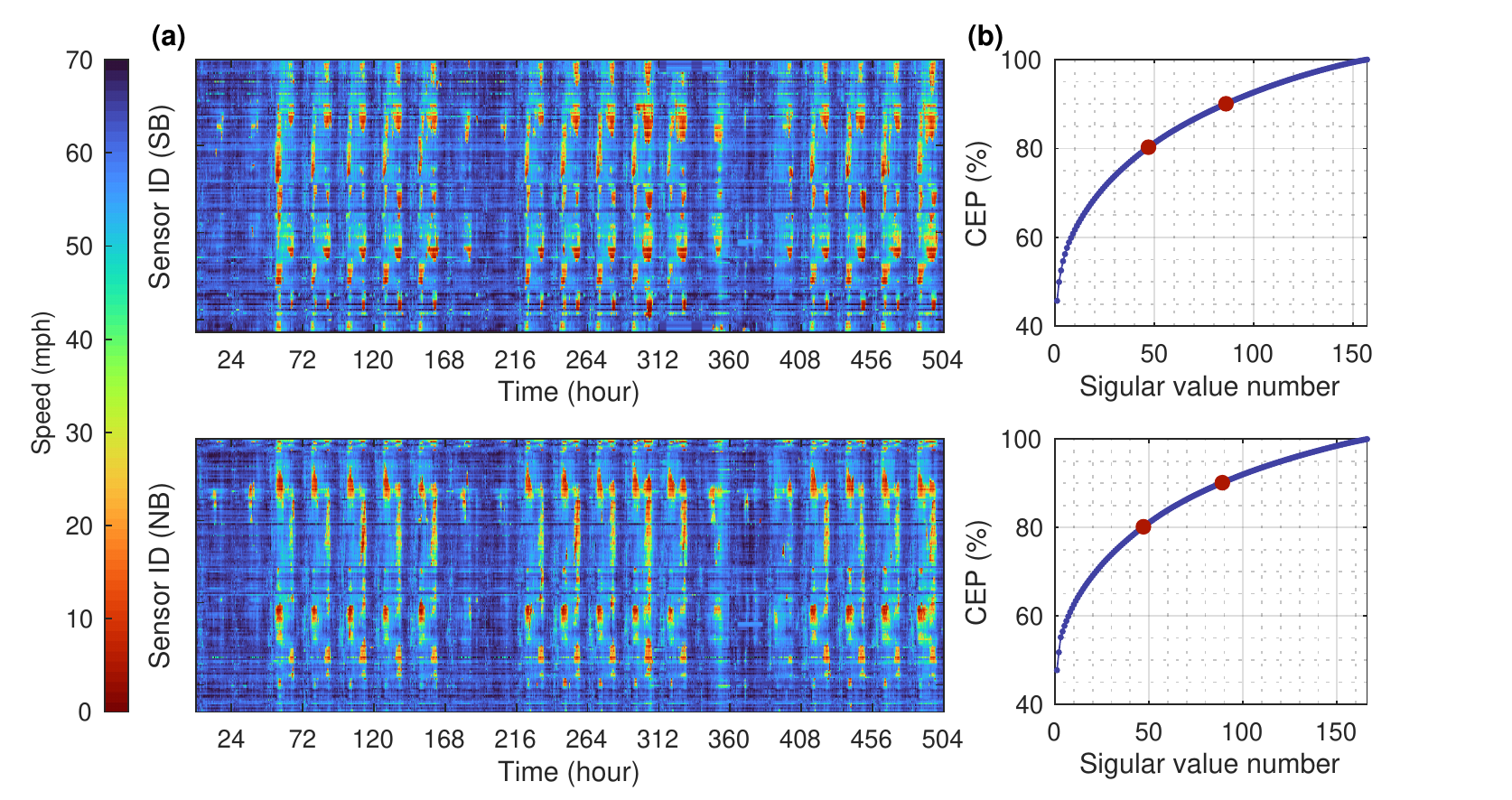}
    \caption{(a): The highway traffic speed data collected from Seattle, USA with 5-minutes resolution; (b): the cumulative eigenvalue percentage (CEP) of training data calculated by singular value decomposition. The red dots represent the first eigenvalue that reaches 80\% and 90\% of CEP, respectively. (top: SB, bottom: NB). }
    \label{fig:data}
\end{figure}

We visualize the training data (the first 336 hours) and the testing data (336 hours to 504 hours) in Figure \ref{fig:data} (a). The visualization clearly showcases the presence of daily/weekly periodicities in both directions of the traffic data. To examine the low-rank characteristic of the data, we depict the cumulative eigenvalue percentage (CEP)---$\sum_{i=1}^ra_i/\sum_{i=1}^{\min(N,T)}a_i$, $a_i$ is the $i$-th largest singular values---in Figure \ref{fig:data} (b). As illustrated, the first 50 singular values account for 80\% of the CEP, indicating that the traffic data can be represented by a few latent factors. The low-rank characteristic can help reduce the complexity of the model, while maintaining good accuracy. The collected data, however, are susceptible to anomalies/noises due to various factors such as car accidents, holidays/events, or sensor malfunctions. For instance, the consecutive missing values for the first several sensors on SB in the training data (between hour 312 and 336) were imputed by copying previous values. Another example is the impact of holidays, such as Martin Luther King Jr. Day, which is a federal holiday in the USA and alters the normal traffic routines. This can be seen in the difference between the first day (Saturday) and the third day (Monday) of the testing data and the corresponding days in previous weeks. The presence of such outliers could significantly impact the performance and application of DMD models \citep{schmid2022dynamic}.

\begin{figure}[!ht]
    \centering
\includegraphics{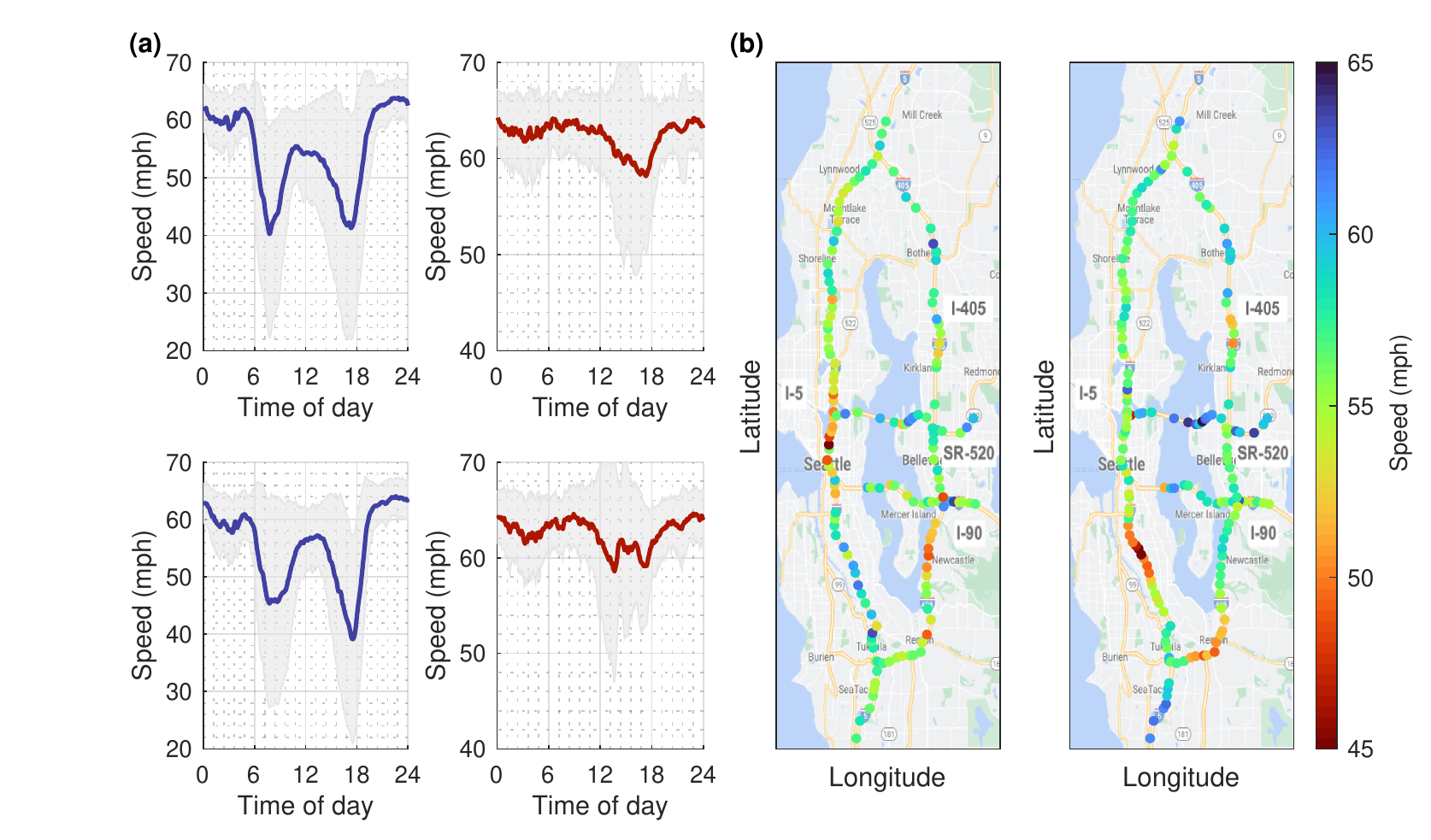}
    \caption{(a): The average traffic speed with standard deviation in SB (top) and NB (bottom). The blue line is for weekdays, and the red line is for weekends, and the gray area denotes the average traffic speed $\pm$ standard deviation;  (b): The average traffic speed of each sensor during the training period in SB (left) and NB (right). The sensors are installed on four highways: I-5, I-90, I-405, and SR-520 (North up).}
    \label{fig:data_st}
\end{figure}

Figure \ref{fig:data_st} shows the average traffic speed profiles of training data from both temporal and spatial perspective. Temporally, the average traffic speed during weekdays follows a typical “M" shape pattern (i.e., with morning/evening rush hours) for both directions during weekdays, while it it appears to be smoother on weekends. Spatially, the average speed at the same location differs in the two directions with considerable heterogeneity. The locations with the lowest average speed (congested areas) are located around the intersections of I-90 and SR-520 in southbound direction, while the congested areas are close to the West Seattle Bridge in northbound direction. In conjunction with Figure \ref{fig:data} (a), we find that the distinct shock waves that typically occur during peak hours, between 6:00 and 10:00, and between 15:00 and 19:00, in proximity to intersections. This phenomenon is likely a result of merging and lane-change maneuvers during high traffic congestion.

\subsection{Traffic speed historical reconstruction and long-term prediction}
\label{sec:reconstruction}

 We first compare the proposed method under different sparsity-promoting parameters $\lambda$ with the following baseline models in historical reconstruction and long-term prediction: 

\vspace{-\topsep}
\begin{itemize}\setlength\itemsep{0em}
    \item DMD: standard Dynamic Mode Decomposition model \citep{kutz2016dynamic}
    \item HDMD: Hankel Dynamic Mode Decomposition \citep{brunton2016extracting, avila2020data}.
    \item fbHDMD: Forward and Backward Hankel Dynamic Mode Decomposition \citep{dawson2016characterizing}
    \item tlsHDMD: Total least square Hankel Dynamic Mode Decomposition \citep{hemati2017biasing}
    \item circDMD (proposed): anti-Circulant Dynamic Mode Decomposition
    \item circDMDsp (proposed):  anti-Circulant Dynamic Mode Decomposition with sparsity-promoting strategy
\end{itemize}

Essentially, these analyzed models are different in three aspects: observables (input), denoising ability, and mode selection. The default DMD operates on the data matrix $\mat{X} \in \mathbb{R}^{N \times T}$, while HDMD, fbHDMD and tlsHDMD use the Hankel matrix $\mat{H} \in \mathbb{R}^{(N\tau) \times (T-\tau+1)}$ defined in \eqref{eq:hankel}, and the proposed circDMD and circDMDsp utilize the anti-circulant matrix $\mat{C}\in \mathbb{R}^{N\tau \times T}$ defined in \eqref{eq:circ} as input. The use of the Hankel matrix and the anti-circulant matrix significantly increases the number of observables, enabling the approximation of the Koopman operator. It is noteworthy that when the delay embedding length is set to $\tau=1$, the Hankel-based and anti-circulant-based DMD models reduce to the standard DMD model. In terms of denoising capacity, forward-backward dynamic mode decomposition (fbDMD) \citep{dawson2016characterizing} and total-least-square dynamic mode decomposition (tlsDMD) \citep{hemati2017biasing} were designed to make the analysis robust to outliers in the data. To evaluate the debiasing performance, we have combined these models with HDMD to produce fbHDMD and tlsHDMD. Finally, the circDMDsp model provides a concise representation by only selecting the dominant modes. In this sense, circDMD is a special case of circDMDsp, where the sparsity-promoting parameter $\gamma=0$. Hence, circDMDsp can be viewed as a general model that balances accuracy, robustness, and meaningful pattern selection.

In the historical reconstruction experiment, we use the optimal hard threshold in Eq.~\eqref{eq:threshold} to determine the rank of the reduced SVD that is used to project the dynamic matrix onto a lower-dimensional subspace for all models. The delay embedding length $\tau$ is set to 3 days ($\tau=3\times 288=864$) for both the Hankel-based models and the anti-circulant-based models. The sparsity parameter $\gamma$, is set to either 500 or 1000 for the circDMDsp model. The reconstruction results are calculated using the spatial modes $\mat{\Phi}$, amplitudes $\mat{b}$, and Vandermonde matrix $\mat{\Psi}$, with $T_\xi=T$ in Eq. \eqref{eq:cdmd} and the matrix inverse operator (e.g., $\mathcal{C}\tau^{-1}(\mat{C}_\text{DMD})$). In the long-term prediction, the spatial modes and amplitudes are kept fixed, but the timestamp length in the Vandermonde matrix is extended to the prediction time window, i.e. $T_\xi = T+288\times7$, for a one-week-ahead prediction.

\begin{table}[!h]
\centering
\small
\begin{tabular}{@{}lllllll@{}}
\toprule
\multirow{2}{*}{Method} & \multicolumn{3}{c}{Southbound direction}                                           & \multicolumn{3}{c}{Northbound direction}                                           \\ \cmidrule(l){2-7}
                        & \multicolumn{1}{c}{Reconstruction} & Pred-F3  & \multicolumn{1}{c}{Pred-L4} & \multicolumn{1}{c}{Reconstruction} & Pred-F3  & \multicolumn{1}{c}{Pred-L4} \\ \cmidrule(r){1-7}
DMD                     & 11.83/14.27                        & 22.38/23.00 & 25.48/25.77                    & 10.24/12.72                        & 19.77/20.46 & 23.82/24.10                    \\
HDMD                    & 2.44/3.71                          & 7.32/10.56  & 4.27/5.36                      & 2.41/3.66                          & 6.55/9.59   & 5.35/6.83                      \\
fbHDMD                  & 8.65/13.27                         & 24.57/31.73 & 37.49/43.19                    & 7.31/11.20                         & 17.43/22.50 & 28.86/34.04                    \\
tlsHDMD                 & 2.41/3.69                          & 6.49/9.59   & 3.24/4.15                      & 2.40/3.64                          & 6.63/9.68   & 5.20/6.75                      \\
circDMD ($\gamma=0$)                & \textbf{2.14/3.20 }                         & \textbf{5.15}/9.45   & 2.76/3.62                      & \textbf{2.12/3.17}                          & \textbf{4.77}/8.39   & 2.46/3.17                      \\
circDMDsp ($\gamma=500$)               & 2.31/3.49                          & 5.15/9.43   & 2.52/3.31                      &  2.24/3.34                        & 4.79/8.40   & 2.41/3.09                      \\
circDMDsp ($\gamma=10^{3}$)               &   3.05/4.67                       & 5.23/\textbf{9.15}   & \textbf{2.51/3.28}                      &   2.94/4.52                        & 4.84/\textbf{8.16}   & \textbf{2.26/2.94}  \\
\bottomrule
\end{tabular}
\caption{reconstruction and prediction experiments: MAE (mph)/ RMSE (mph) of baseline models and the proposed models under $\tau=3\times288$. }
\label{tab:reconstruction}
\end{table}

Table \ref{tab:reconstruction} shows the resulting MAE and the RMSE for traffic speed reconstruction and prediction. The prediction accuracy is evaluated for two different scenarios, including the first three days with a holiday, denoted as Pred-F3, and the remaining four days, denoted as Pred-L4. The standard DMD fails to estimate the traffic speed with the highest MAE and RMSE in all tasks. It is because the observables of DMD are limited to capturing the whole traffic dynamics. Generally, Hankel-based DMDs and circulant-based DMDs outperform the standard DMD model, with the exception of fbHDMD. The fbHDMD is under the assumption that the dynamics of the original system are invertible \citep{dawson2016characterizing}; however, $\mat{\Tilde{A}}_f$ and $\mat{\Tilde{A}}_b^{-1}$ are typically projected onto different subspaces in the transportation network. Another denoised DMD, tlsDMD, which is based on total least squares to address noise, shows marginal improvement in the accuracy compared to HDMD in SB. However, it exacerbates the performance in NB. The proposed circDMD and circDMDsp demonstrate the best performance in both reconstruction and prediction tasks. As an extension of circDMD, circDMDsp applies a sparse-promoting strategy to reduce the rank and optimize the amplitudes of modes, resulting in a more parsimonious representation. The reconstruction performance decreases as the value of $\gamma$ increases, as fewer spatiotemporal patterns are used in the model. However, the prediction performance slightly improves with a larger value of $\gamma$. It reflects that a few dominant patterns capture the main traffic speed characteristics.Based on the performance in both tasks, we set $\gamma=500$ in circDMDsp to analysis in the following sections.

\begin{figure}[htpb]
    \centering    \includegraphics{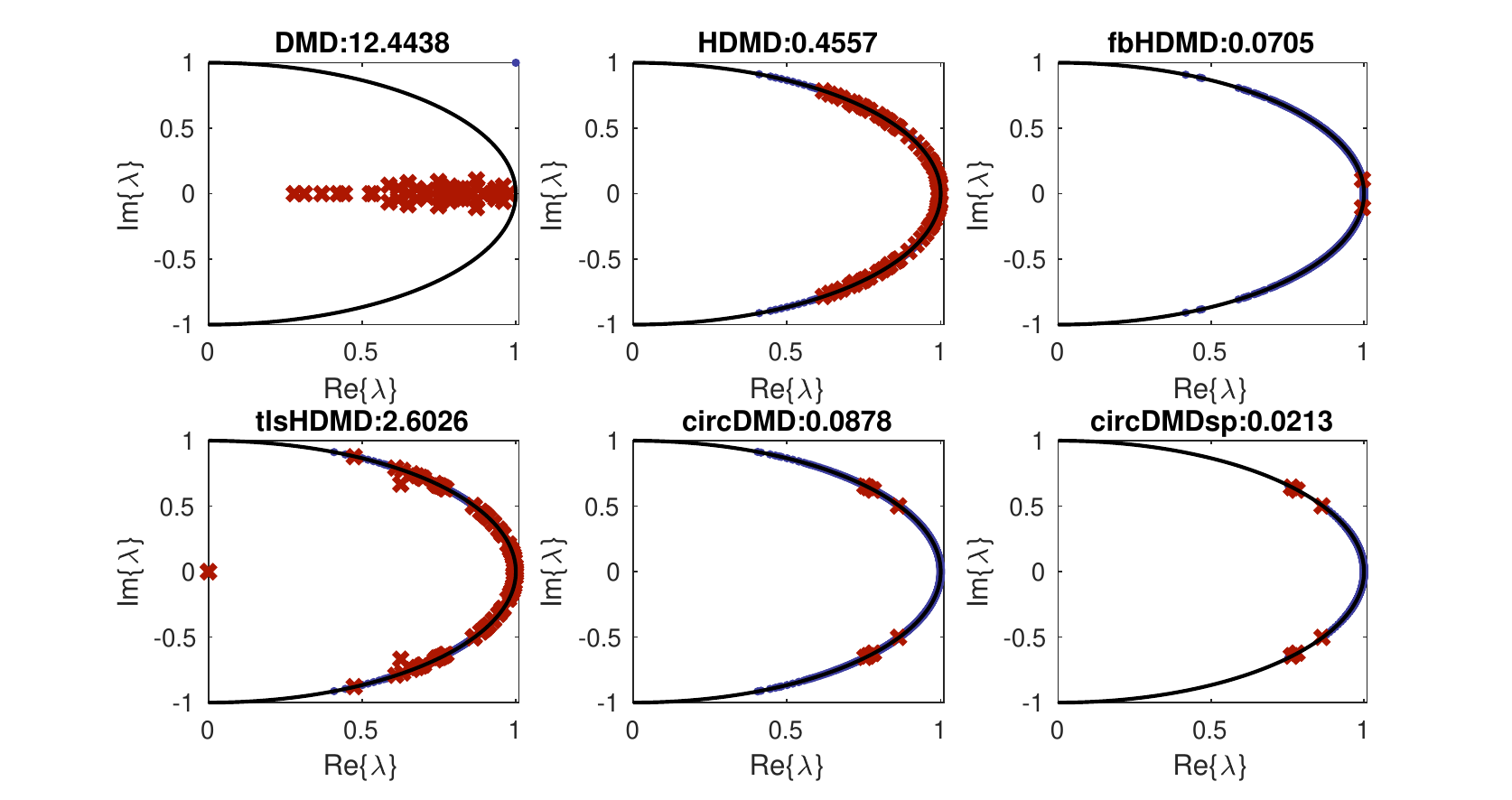}
    \caption{The eigenvalues $\mat{\lambda}$ with the unit circle on the complex plane in SB. The dot in blue denotes steady dynamic modes and the cross in red denotes varying dynamic modes. The sum of absolute differences between $|\mat{\lambda}|$ and $\mat{1}$ is shown in the title. }
    \label{fig:eigenvalue}
\end{figure}

The estimation performance is dependent upon the dynamic modes $\mat{\Phi}$ (eigenvectors of the dynamic matrix) and the temporal dynamics calculated by $\mat{\lambda}$ (the corresponding eigenvalues of the dynamic matrix). {Specifically, the magnitude of the eigenvalue $|\lambda_i|$ determines whether the corresponding dynamic mode is steady ($|\lambda_i|=1$) or varying (decaying $|\lambda_i|<1$, growing $|\lambda_i|>1$). Namely, a decaying dynamic mode will ultimately converge to zero, and a growing dynamic mode will diverge to infinite. Both of them would negatively impact the long-term prediction performance. On the contrary, a steady dynamic mode is invariant or slightly variant to help predict future values.} Figure \ref{fig:eigenvalue} provides the eigenvalues obtained from the DMD-based models in SB by plotting them on the complex plane with the unit circle. The steady eigenvalues, which are defined as $(1 - 10^{-3})\leq |\lambda_i| \leq( 1 + 10^{-3})$, are colored blue, while the varying eigenvalues beyond this range are colored red. The sum of absolute differences between $|\mat{\lambda}|$ and $\mat{1}$ is also displayed in the title of each subfigure. Figure \ref{fig:eigenvalue} demonstrates that all of the eigenvalues are within the unit circle in DMD,  resulting in the rapid decay of the corresponding spatial patterns and poor estimation performance. In contrast, most eigenvalues of other methods are located around the circle unit, especially the proposed circDMDsp model has the minimal absolute summation.

\subsection{Spatiotemporal dynamic pattern analysis}
\label{sec:pattern}

The unique advantage of DMD-based models is their ability to produce interpretable results through eigendecomposition of the dynamic matrix. We analyze the spatiotemporal patterns from a dynamic view by the proposed circDMDsp in this subsection.

\begin{figure}[!ht]
    \centering
    \includegraphics{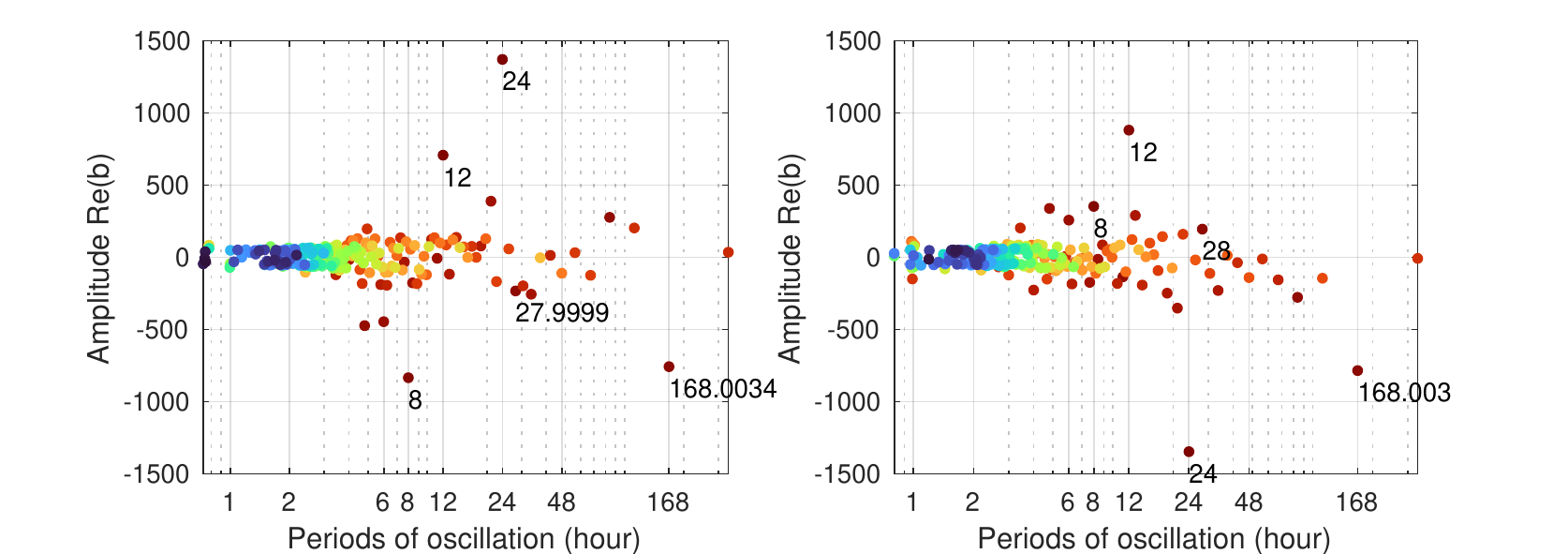}
    \caption{{Periods of oscillation with amplitudes $\mat{b}$ in SB (left), and in NB (right). Each dot represents a period of oscillation (hour). The infinite period ($Re{(b)}=-2.099\times 10^4$) and negative periods are excluded in the figure.}}
    \label{fig:period}
\end{figure}

First, the phase of eigenvalues $\angle \lambda_i$ determines the oscillation period of the dynamic modes \citep{brunton2016extracting}:
\begin{equation}
    \text{period}_i = \frac{2\pi\Delta t}{\imag(\log(\lambda_i))},
\end{equation}
where $\imag()$ is the imaginary part of eigenvalues and sampling time $\Delta t = 1/12$ hour (5 minutes) in the study. { The periods of oscillation, along with their corresponding amplitudes $\mat{b}$, are depicted in Figure \ref{fig:period}. Oscillation periods with infinite duration ($\imag(\lambda_i)=0$) and negative periods (conjugate eigenvalues of positive periods) have been excluded. The figure suggests that longer oscillation periods are typically associated with larger amplitude. In other words, the traffic speed is mainly controlled by low-frequency patterns. The infinite period has the largest amplitude of $Re{(b)}=-2.099\times 10^4$, revealing that the non-periodic dynamic mode is the most dominant pattern in the data. The second most dominant pattern has a period of 24 hours, confirming the daily similarity in traffic speed. The weekly pattern (168 hours) is also detected from the model. In addition to the daily and weekly patterns, intra-day patterns (with period of 12 hours, 8 hours, 6 hours, 4.8 hours) and intra-week patterns (with period of 84 hours, 28 hours, and 11.2 hours) have also been identified as dominant patterns. All of these periods are discovered from the data without any prior information and reflect the inherent evolution of traffic speed. They can be utilized as traffic features to address downstream tasks, such as the long-term traffic prediction mentioned in the previous section. }

\begin{figure}[t]
    \centering
    \includegraphics{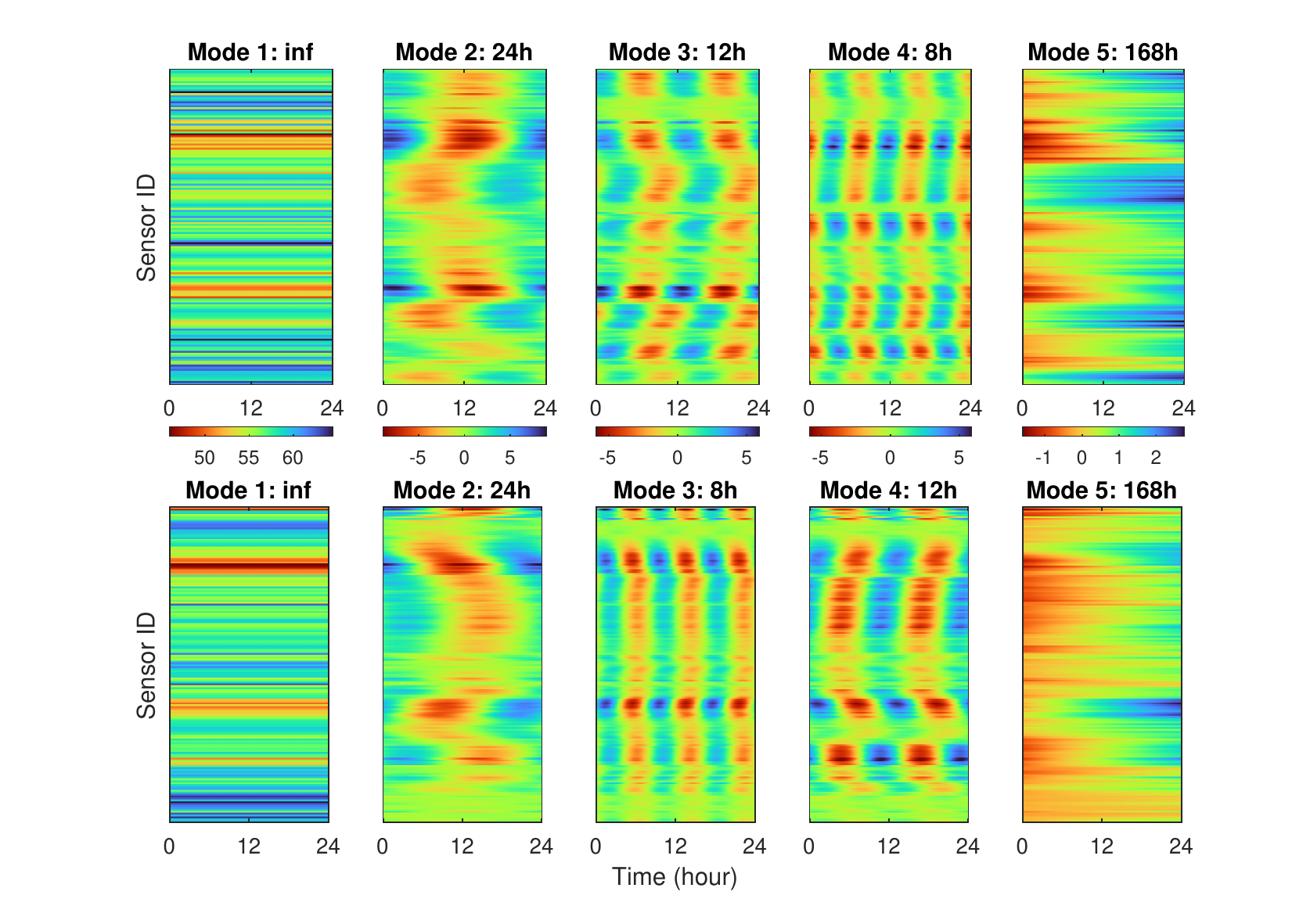}
    \caption{The real part of first five reshaped dynamic modes $\mat{\phi}_ib_i \in \mathbb{R}^{N\times \tau}, i=1,\dots,5$ with corresponding periods shown in title (top row: SB, bottom row: NB). }
    \label{fig:mode}
\end{figure}

The spatial modes $\mat{\Phi}$ capture the spatial coherence among sensors. We next examine how the spatial modes evolve with their corresponding amplitudes $\mat{b}$ and oscillation periods. To achieve this purpose, we define a reshape operator--- $\operatorname{reshape}(\mat{\phi}_ib_i,N,\tau)$---to organize every spatial mode $\mat{\phi}_i \in \mathbb{R}^{N\tau}~(i=1,\cdots, r)$ with amplitudes to a matrix of size $N \times \tau$. As we set the delay embedding length to $K=288\times3$ (3 days) in this study, the dynamic modes naturally represent how the spatial patterns evolve over 3 days. {Figure \ref{fig:mode} shows how the five most significant modes (out of the 441 modes identified by circDMDsp) evolve in the first 24 hours, and the other first ten modes are shown in Figure \ref{fig:southbound} and Figure \ref{fig:northbound} in Appendix. We can find that the traffic speed data combines a series of periodic patterns with different amplitudes/weights. Remarkably, the constant/mean pattern (Mode 1), the daily pattern (Mode 2), and the weekly pattern (Mode 5) are three dominant spatiotemporal structures detected by circDMDsp. Unlike other matrix factorization methods, which solely separate spatial patterns and temporal patterns, the proposed model finds coherent spatiotemporal patterns because of the locally linear assumption $\mat{x}_{t+1} = \mat{A}\mat{x}_t$. }

 \begin{figure}[t]
    \centering
    \includegraphics{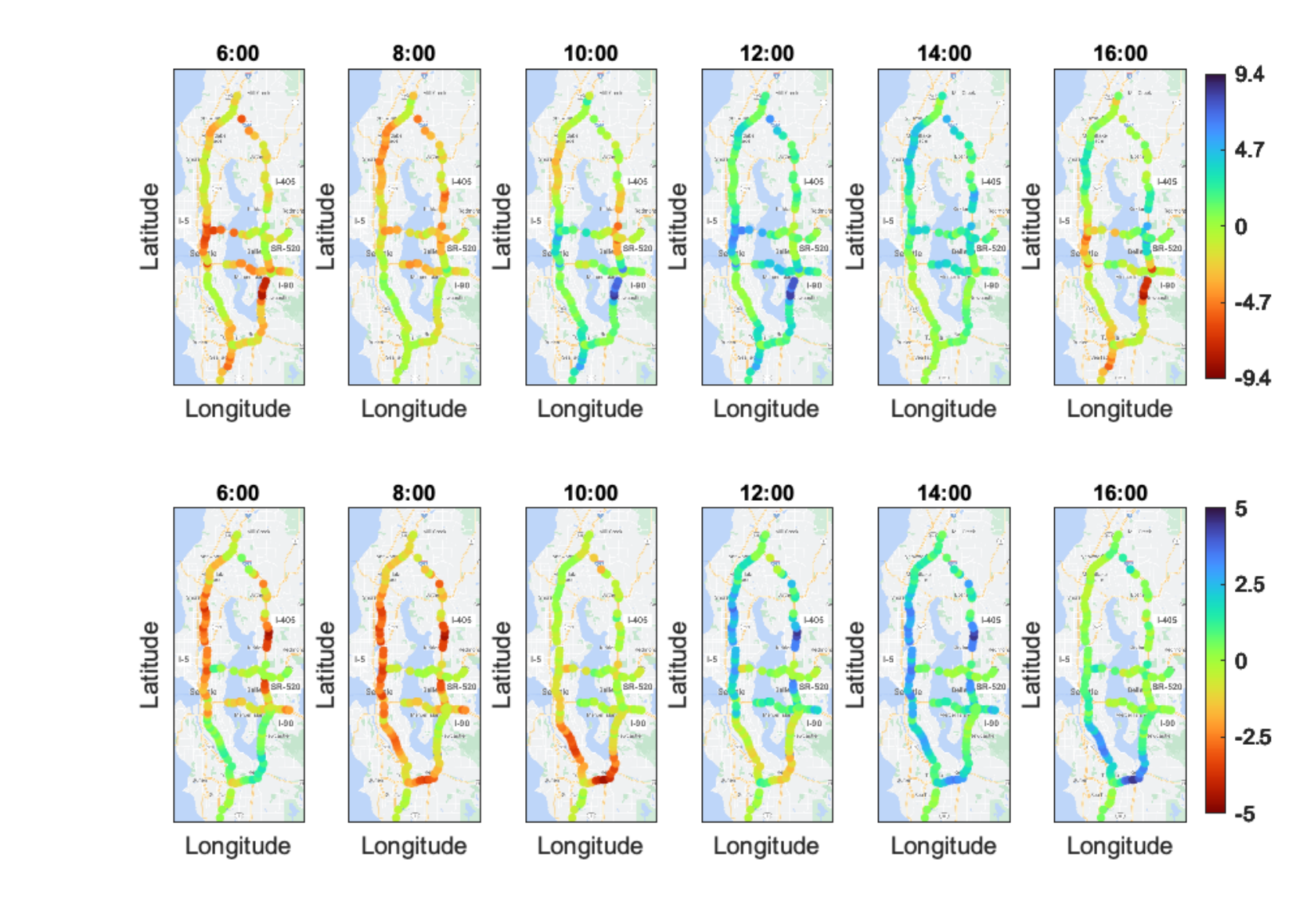}
    \caption{The Mode with a period of 12 hours from 6:00 to 16:00. The title of each subplot is the timestamp. Each dot represents a sensor's location, and the color denotes the magnitude of the sensor (top row: SB, bottom row: NB).}
    \label{fig:12h}
\end{figure}

{For better visualization, we map the mode with a 12-hour period (Mode 3 in SB, Mode 4 in NB) from 6:00 to 16:00 at 2-hour intervals on Seattle highways in Figure \ref{fig:12h}, and visualize the obtained spatial structure at different times of a day. We can detect congestion spots within a 12-hour period. Congested spots refer to the nearby sensors with negative speed---decreased the aggregated speed from all of the modes. For example, some congested spots are found at 6:00 to 8:00 and 16:00 in southbound traffic. They are located near intersections, e.g., between I-5 and SR-520 or I-90 and I-405. The situation is relieved (turns to be positive speed) in the early afternoon. The location and time of congestion spots are different in northbound traffic, where the congestion spots are more likely to locate at I-5 and I-90 intersection or I-405 and SR-520 intersection. In addition, more sensors are regarded as congestion spots in I-5 at 8:00 in NB. The heterogeneous distribution of dynamic modes indicates that traffic characteristics are impacted by traffic flow direction. }

\begin{figure}[!htbp]
    \centering
    \includegraphics[width=\linewidth]{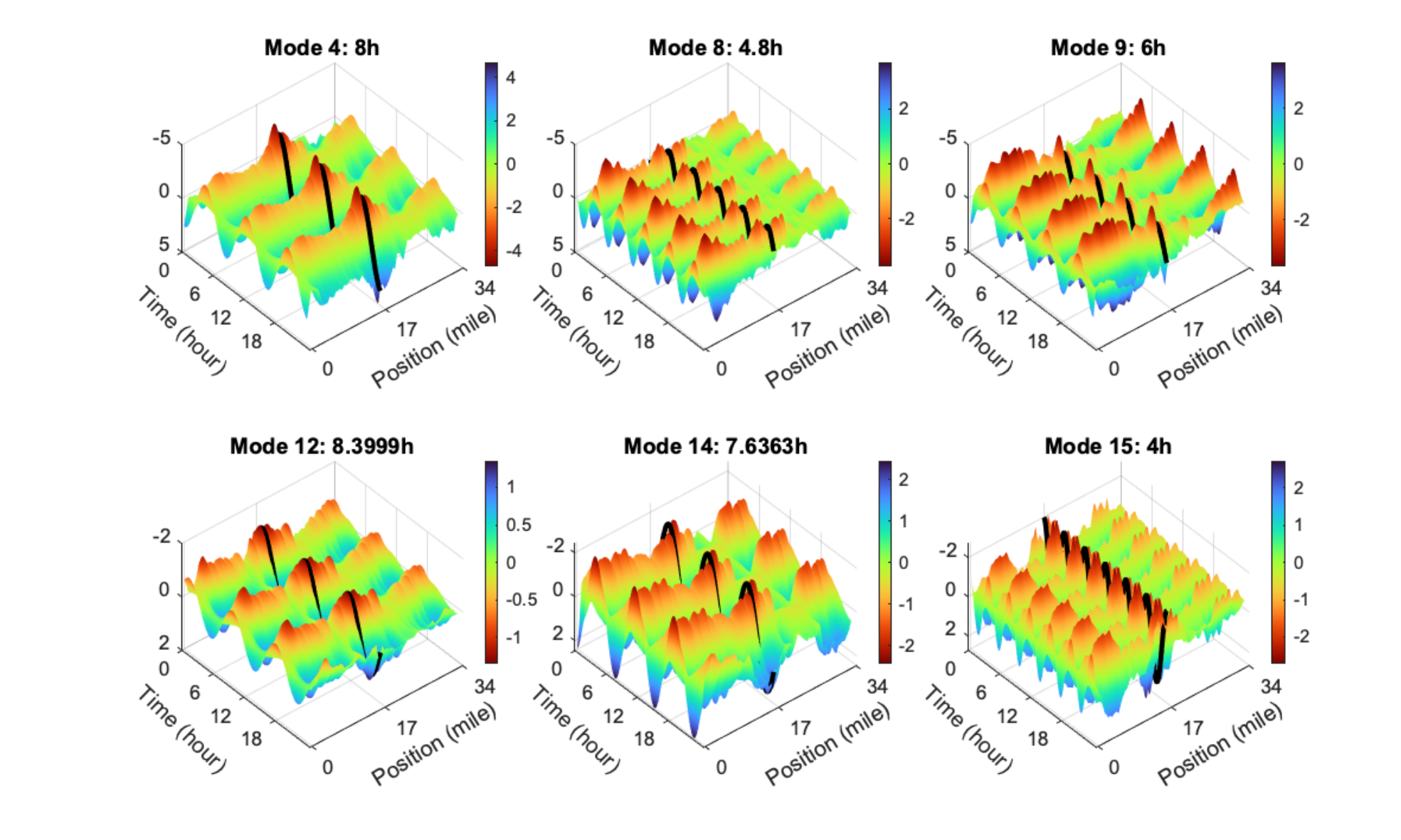}
    \caption{The 24 hours evolution of 6 dynamic modes with corresponding amplitudes along I-5 highway in SB. The intersection between I-5 and SR-520 is labeled with black line. The z-axis is reversed.}
    \label{fig:i5}
\end{figure}

{The dynamic modes can also be analyzed for a specific highway segment. For example, Figure \ref{fig:i5} displays the 24-hour evolution of 6 selected dynamic modes along the I-5 highway in southbound traffic. The sensor installed at the intersection between I-5 and SR-520 is labeled with a black line. The hidden spatiotemporal wave structures are uncovered from the model. The results suggest that an apparently localized cluster is just before the intersection in intra-day patterns (e.g., Mode 4, Mode 8, Mode 9, and Mode 15). These four modes are different in other positions along the highway. For example, Mode 9 shows a peak at the end of the highway, while Mode 8 and Mode 15 are at the beginning of the highway. It might be impacted by other intersections on I-5, which the information is not included in the study. Mode 12 and Mode 14 are intra-week patterns, where the localized cluster occurs after the intersection.}

\subsection{Short-term predictability of traffic speed}
\label{sec:predict}

{Section \ref{sec:pattern} suggests that the traffic speed can be decomposed into a set of periodic spatiotemporal patterns. Long-term traffic patterns are generally related to seasonal trends and regularities, while short-term traffic patterns are associated with non-seasonal features \citep{boukerche2020machine}. The proposed circDMDsp is good at long-term prediction (e.g., daily or weekly prediction) because the spatiotemporal patterns are periodic functions (Figure \ref{fig:mode}), capturing global consistency. On the contrary, the prediction performance from circDMDsp could be unsatisfying in some scenarios where short-term traffic factors play a vital role in prediction, such as uncertain fluctuations caused by traffic maneuvers. We next use short-term predictability to evaluate to what extent the data from a sensor can be approximated by DMD. We use MAPE conducted from cicrDMDsp to evaluate such short-term predictability at the sensor level.}

\begin{figure}[!ht]
    \centering
\includegraphics{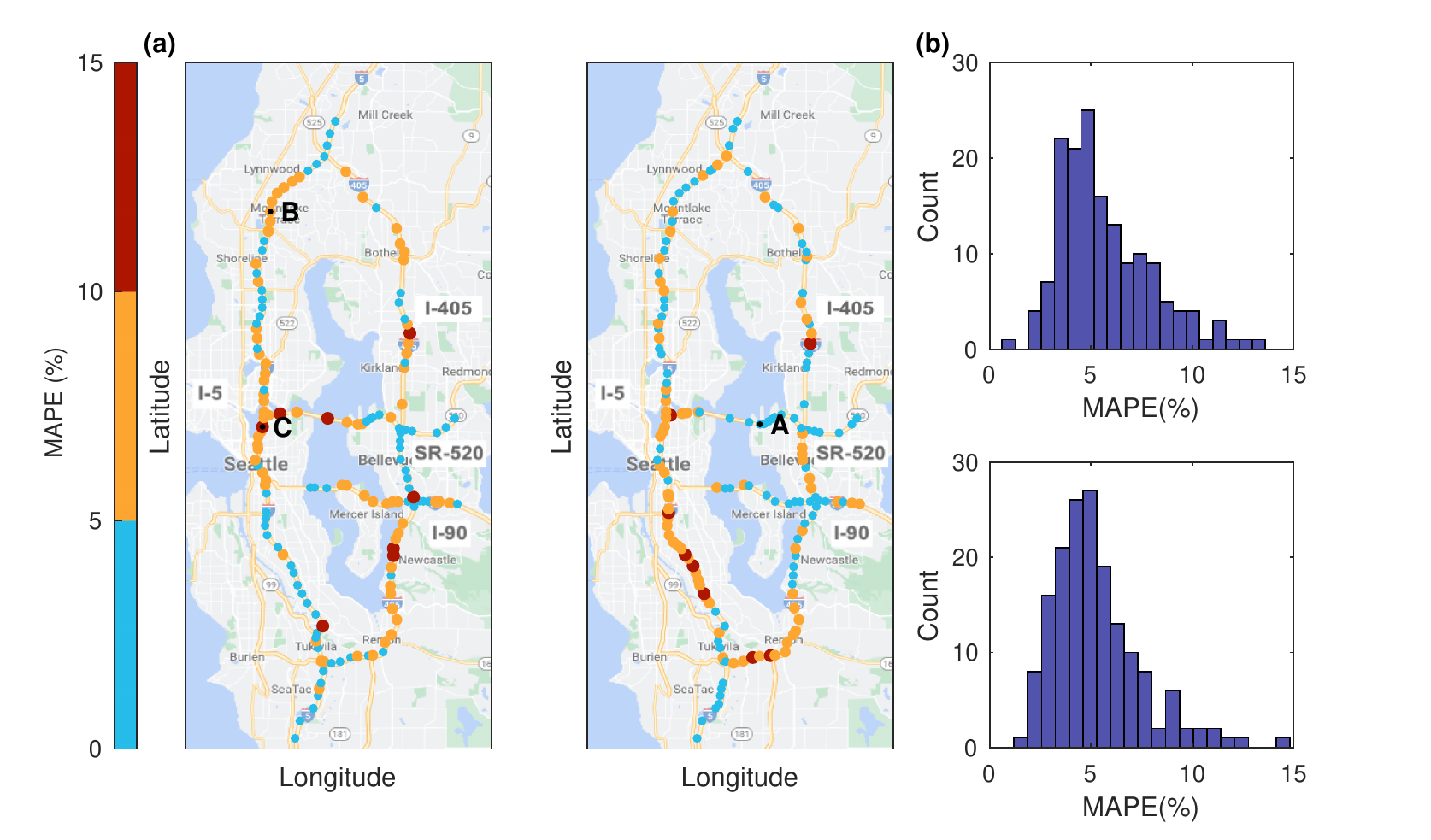}
    \caption{(a): The groups of short-term predictability in terms of MAPE in SB (left) and NB (right); (b): The histogram of MAPE in SB (top) and NB (bottom). The location of SensorA, SensorB, and SensorC are labeled on the map. (North up)}
    \label{fig:sensor_pre}
\end{figure}

Figure \ref{fig:sensor_pre} (a) maps the MAPE of historical reconstruction from Section \ref{sec:reconstruction} on Seattle highways (I-5, I-90, I-405, and SR-520). Figure \ref{fig:sensor_pre} (b) shows the histogram of MAPE, which suggests that the MAPE of most sensors are around 5\%, and only a few of them are larger than 10\% in both directions. We separate all sensors into three groups in terms of their MAPE. The sensors with smaller MAPE can be explained better by the model; this also suggests that these sensors are basically dominated by periodic and recurrent patterns, and the proposed model is sufficient to achieve good prediction accuracy. On the other hand, those sensors with higher MAPE are located at the traffic headed to Seattle or the intersections of highways with strong non-recurrent patterns. These locations usually have higher traffic flow and more complicated traffic maneuvers, which need more local information to perform prediction. Short-term prediction for these sensors with strong non-recurrent patterns would require a better model---such as deep learning-based time series models---to achieve better prediction performance.

\begin{figure}[!ht]
    \centering
    \includegraphics{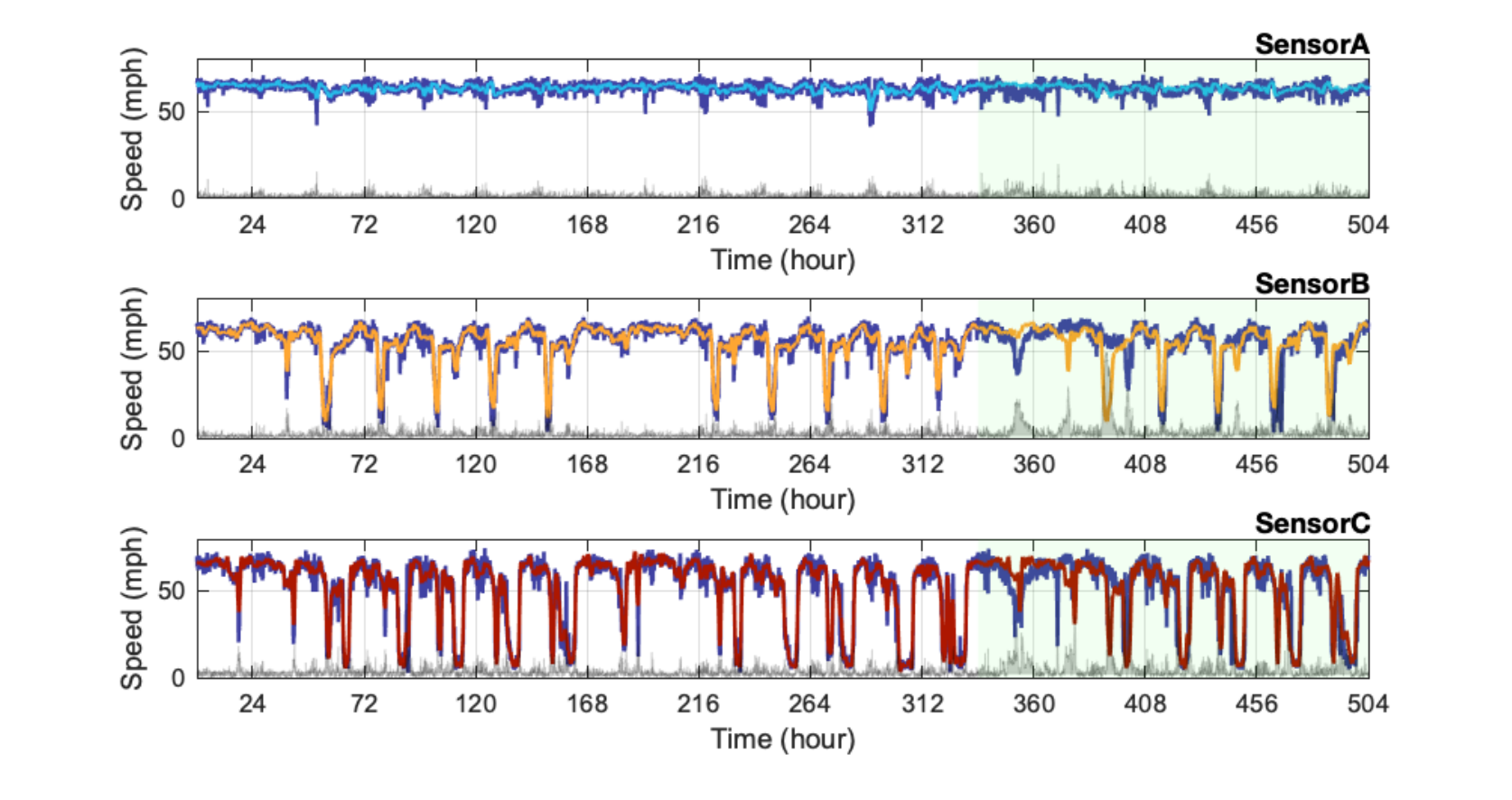}
    \caption{The 14-day estimation and 7-day prediction (green background) results by circDMDsp model of three selected sensors. Purple line: ground-truth of each sensor; blue, orange, and red line: estimation of each sensor; gray line: absolute residuals of each sensor. The location of SensorA, SensorB, and SensorC are labeled in Figure \ref{fig:sensor_pre} (a).} 
    \label{fig:prediction}
\end{figure}

We next select three sensors in each category and evaluate their predictability. The 14-day historical reconstruction and the 7-day prediction (in the green background) of the three selected sensors by circDMDsp are shown in Figure \ref{fig:prediction}. SensorA has the smallest MAPE among these three sensors. It can be found that the traffic speed collected by SensorA evolves regularly---the speed data is almost constant, and its temporal variation is smooth without any abrupt changes. The prediction results of SensorA are accurate as the traffic characteristics can be captured well by the discovered spatiotemporal patterns. For SensorB, the overall MAPE is above 5\%; however, we can see that the prediction is  very good over the last four days, and the unsatisfying performance for the first three days is mainly due to that the third day (Monday) of the prediction period is a holiday (Martin Luther King Jr. Day), and the traffic routines are different from the previous Monday in the training periods. For example, there would be traffic congestion in the morning peak on Mondays, but this phenomenon is nonexistent in the prediction stage due to the holiday. The MAPE for SensorC is more than 10\%. The challenge in predicting SensorC is that the sensor is located at the intersection between I-5 and SR-520, and it contains many non-recurrent sharp transitions between free flow and severe traffic congestion. This type of transition is mostly dominated by short-term information, and it is difficult to match the exact peak hour based on periodicity alone. For this type of sensor, we suggest using more advanced short-term prediction frameworks such as deep learning-based time series models.

\begin{figure}[!t]
    \centering
    \includegraphics{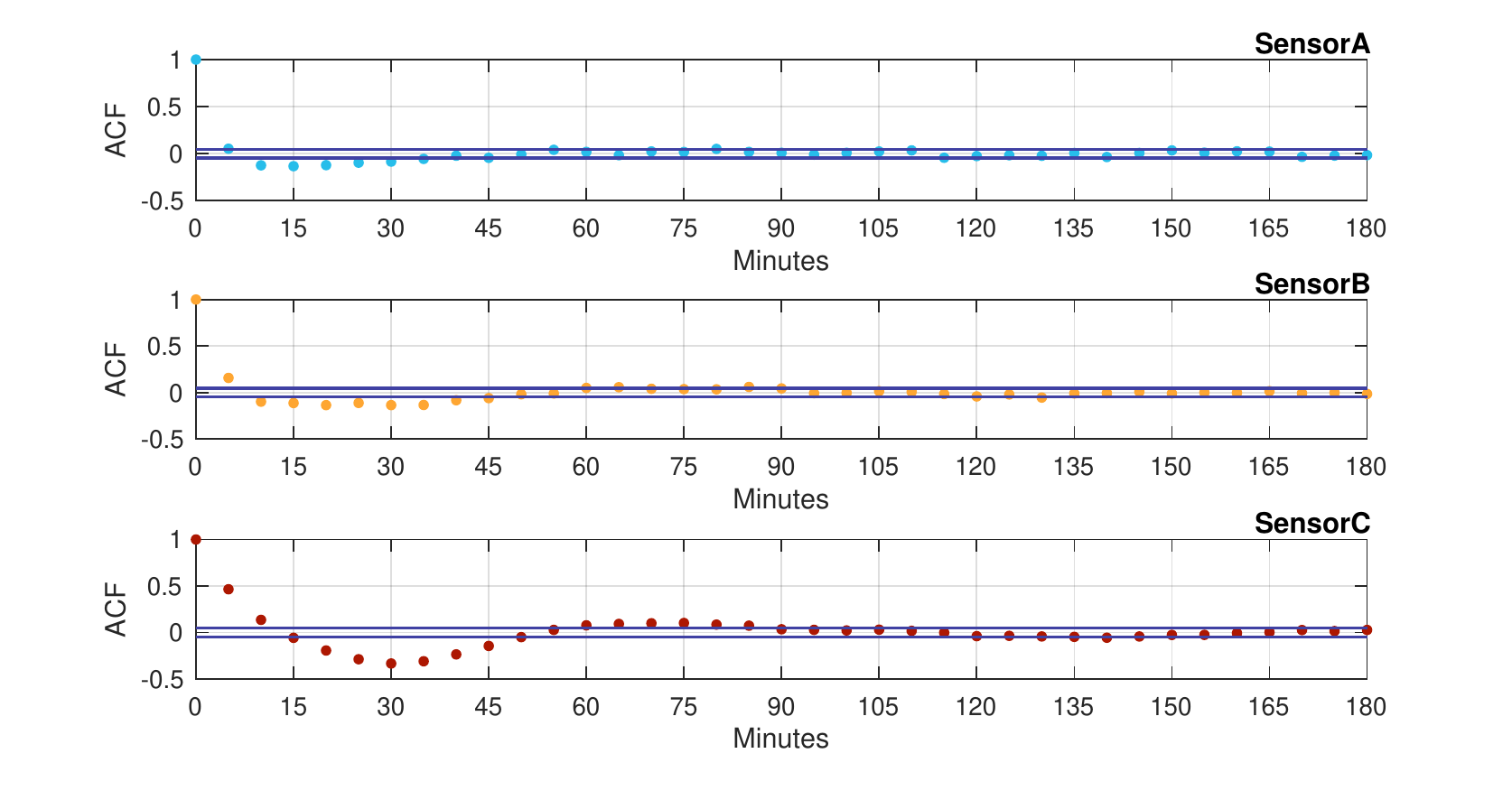}
    \caption{The autocorrelation function (ACF) of historical reconstruction residuals from SensorA, SensorB and SensorC with 5-min time lag. Blue lines are the 3 times standard errors in the confidence bounds.  }
    \label{fig:ACF}
\end{figure}

To further clarify the short-term predictability of sensors, we analyze the autocorrelation function (ACF) of the residual process (gray line in Figure \ref{fig:prediction}) of historical reconstruction. ACF reveals self-similarity over different time lags. The coefficients of each time lag represent the correlation. Figure \ref{fig:ACF} shows the ACF of SensorA, B, and C with three times standard errors in the confidence bounds (blue line), respectively. SensorA shows very weak autocorrelation, suggesting that circDMDsp itself has already achieved very good reconstruction, and there is not much to be learned from the residual process. For SensorB and SensorC, the autocorrelation reaches 0 when the time lags are around 30 min and 60 min, respectively. This result suggests that reconstruction/prediction for SensorB/SensorC can indeed benefit from a better time series model that can effectively characterize the residual processes (with properly selected lags).

\begin{figure}[!t]
    \centering
    \includegraphics{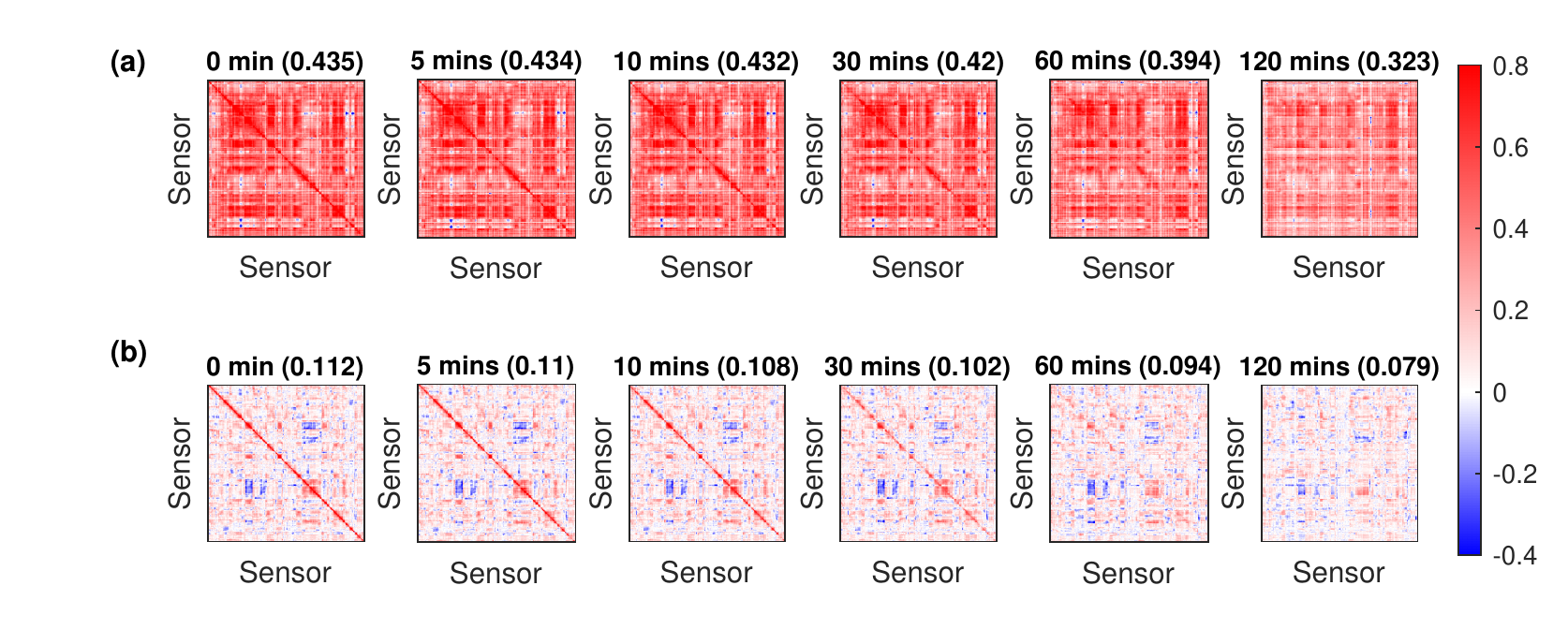}
    \caption{Residual correlation matrices of Pred-F3 (a) and Pred-L4 (b) at different time lag $l$. The average of absolute values of correlations are shown in the parentheses.  }
    \label{fig:corr}
\end{figure}

We can further analyze the spatiotemporal correlation structure of the residual process. In doing so, we calculate the correlation matrix of prediction residuals at different time lags to analyze the short-term predictability at the network level. Let $\mat{\eta}_t \in \mathbb{R}^{N}$ be the residuals at timestamp $t$, then the correlation matrix between $l$ time lag is defined as $\operatorname{corr}(\mat{\eta}_{t-l}, \mat{\eta}_t)$. To exclude the impact of holidays in the analysis, we perform separate correlation analyses for ``Pred-F3'' (3 days) and ``Pred-L4'' (4 days), respectively. Figure \ref{fig:corr} exhibits the residual correlation matrices of Pred-F3 and Pred-L4 at time lags $l = 1,2,6,12$, corresponding to $5,10,30,60,120$ mins; sensors are in sequential order based on their locations on the highway. Figure \ref{fig:corr} (a) shows the correlation structure for the first three days, and we observe strong positive correlations across all sensor pairs. However, this is a simple artifact resulting from the ``systematic but unpredictable'' traffic during the holidays. We focus more on the analysis for the last four days as presented in  Figure \ref{fig:corr} (b). As can be seen, we do observe a strong diagonal structure when time lags are small, suggesting that the residual processes are spatially correlated. Overall, the  correlations weaken with the increase of time lags and almost become 0 when $l=12$ (i.e., 120 min). This further confirms that, for the analyzed Seattle data, short-term traffic prediction is only meaningful when the prediction window is less than 2 hours; otherwise, one should rely on long-term prediction frameworks that leverage the periodic, seasonal and recurrent patterns of the data. We believe this finding should be universal for any traffic time series prediction tasks.

The short-term predictability analysis can also help ITS reveal and rank sensor importance in real-world operations. For example, sensors/segments with good short-term predictability are mainly dominated by the periodic patterns learned from the proposed circDMDsp model, and thus it is simple to perform prediction and interpolation for these sensors/segments. On the contrary, sensors/segments with unsatisfying short-term predictability often exhibit non-recurrent patterns and short-term variations caused by specific events/incidents; in these cases, it becomes more important to incorporate local/side information for real-time predictions and management.

\section{Conclusion and future work}
\label{sec:conclusion}

In this paper, we propose a novel DMD framework---anti-Circulant Dynamic Mode Decomposition with sparsity-promoting (circDMDsp)---to analyze highway traffic speed data from a dynamic systems perspective. The proposed circDMDsp framework decomposes the traffic speed data into dynamic modes, temporal dynamics and corresponding amplitudes, allowing us to perform both reconstruction/denoising of the input data and long-term prediction/extrapolation for future data. Overall, the advantages of the proposed model are that it does not require large training data and there exist only a few hyper-parameters to train the model. Assuming that circDMDsp can fully learn and capture the recurrent patterns in the data, we also examine the predictability of highway traffic speed data. Our results on the Seattle highway data suggest that short-term traffic prediction tasks are meaningful if the prediction window is less than an hour, which we consider corresponding to the general temporal impact range of nonrecurrent events; beyond that, there is no short-term information to learn and long-term prediction can be achieved with high accuracy with a simple estimation---such as historical average and the proposed circDMD---given the strong periodic, recurrent and rhythmic nature of traffic data.

{There are several directions for future research. Firstly, the resulting spatiotemporal dynamic patterns from the proposed framework can be used in many downstream applications/tasks. For example, the dynamic modes of sensors $\mat{\Phi}$ can be used to study the dynamic similarity \citep{wang2022extracting}, and the temporal evolution $\mat{\Psi}$ can be used as time embedding to support deep learning-based forecasting models. Secondly, the results and analyses presented in Section~\ref{sec:predict} can be used to guide short-term traffic prediction tasks. For instance, by using circDMDsp as a surrogate model to remove long-range seasonality/periodicity, we can design better short-term prediction models with a focus on better predicting those nonrecurrent phenomena. The sensors with the smaller MAPE could be well predicted by regular patterns discovered by circDMDsp, but other sensors should incorporate more information or apply other models (e.g., deep learning models) to achieve better prediction performance. Thirdly, the proposed framework can be further extended to an online version---updating the model with the arrival of new data \citep[see e.g., ][]{cheng2022real}, making both short-term and long-term predictions more accurate. Lastly, a promising research direction is to develop DMD-based methods that can directly work with corrupted traffic data with missing values, by leveraging the low-rank properties of both the observations \citep[see, e.g., in][]{chen2021bayesian,lei2022bayesian} and their dynamics  \citep[see, e.g., in][]{chen2022discovering}.}

\section*{Acknowledgment}

This research is supported by the Natural Sciences and Engineering Research Council (NSERC) of Canada and the Canada Foundation for Innovation (CFI). X. Wang would like to thank FRQNT for providing the B2X Doctoral Scholarship.

\bibliographystyle{elsarticle-harv}
\bibliography{ref}

\newpage
\section*{Appendix}
\begin{figure}[!htbp]
    \centering
    \includegraphics[width=0.75\linewidth]{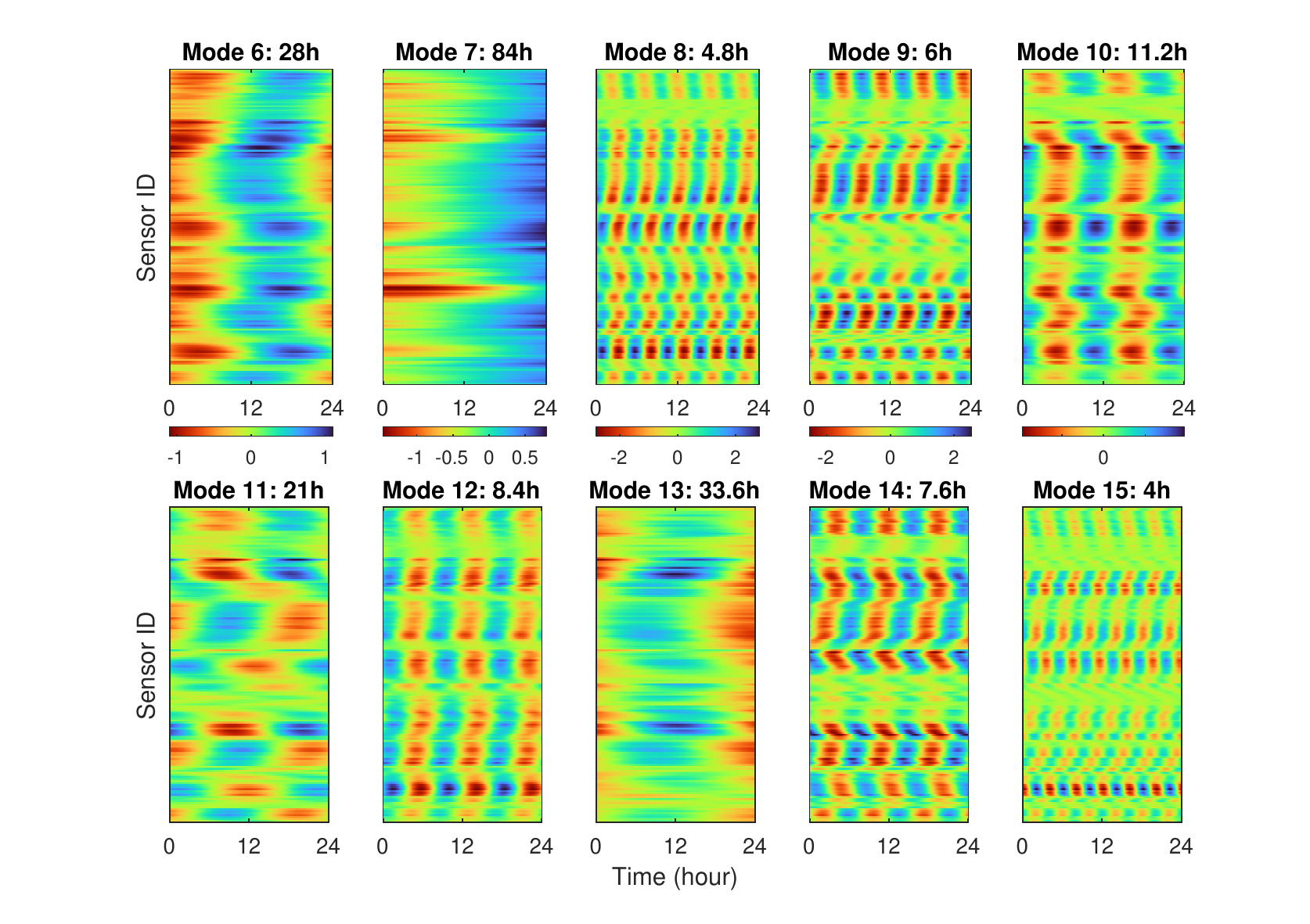}
    \caption{The real part of first 6 to 15 reshaped dynamic modes $\mat{\phi}_i b_i \in \mathbb{R}^{N\times \tau}, i=6,\dots,15$ with corresponding periods in southbound traffic.}
    \label{fig:southbound}
\end{figure}

\begin{figure}[!b]
    \centering
    \includegraphics[width=0.75\linewidth]{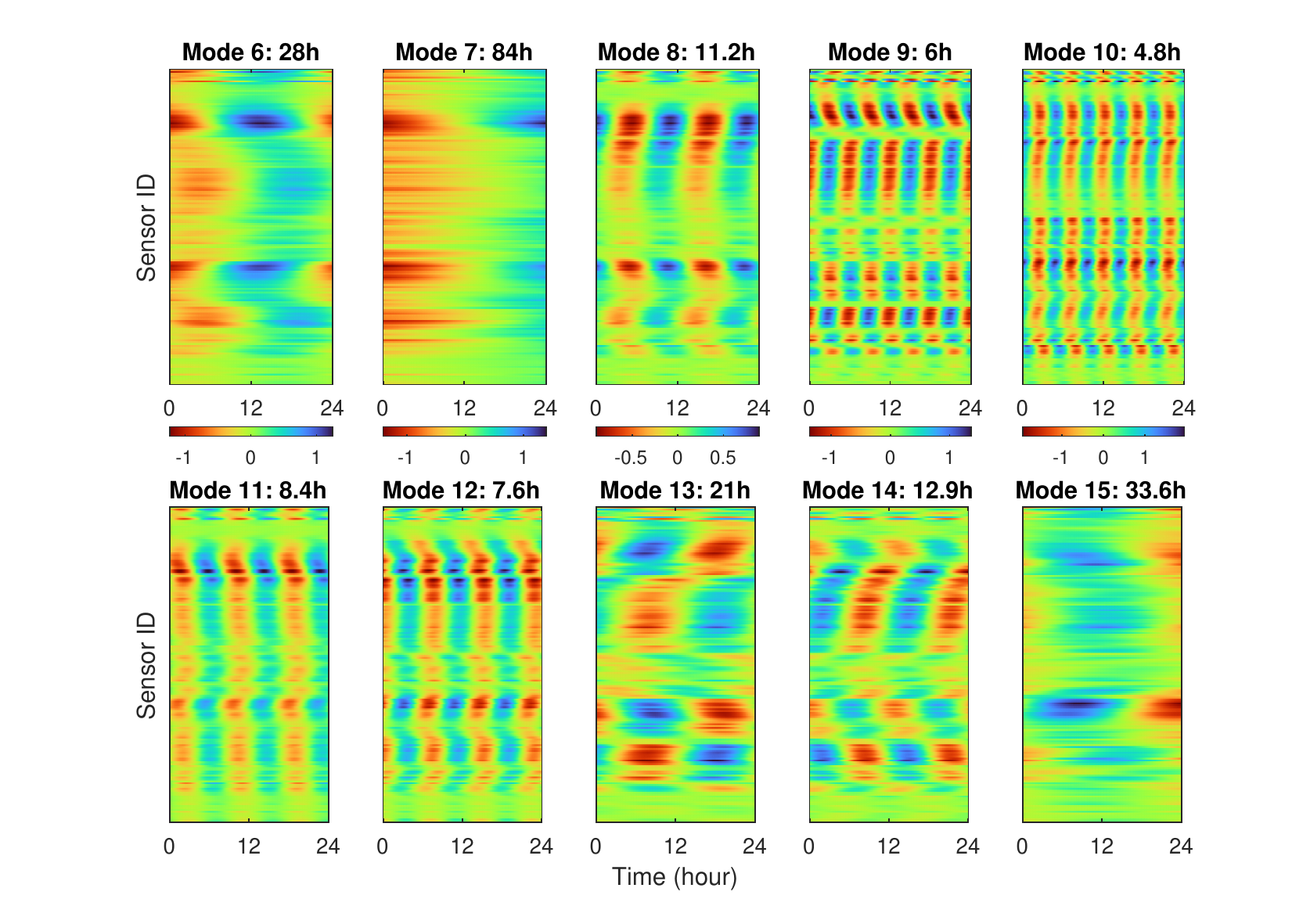}
    \caption{The real part of first 6 to 15 reshaped dynamic modes $\mat{\phi}_i b_i\in \mathbb{R}^{N\times \tau}, i=6,\dots,15$ with corresponding periods in northbound traffic.}
    \label{fig:northbound}
\end{figure}

\end{document}